\documentstyle[12pt,psfig] {article}

\newcommand{\be}{\begin{equation}}
\newcommand{\ee}{\end{equation}}
\newcommand{\bd}{\begin{displaymath}}
\newcommand{\ed}{\end{displaymath}}
\newcommand{\ba}{\begin{array}}
\newcommand{\ea}{\end{array}}
\newcommand{\bt}{\begin{tabular}}
\newcommand{\et}{\end{tabular}}
\newcommand{\bc}{\begin{center}}
\newcommand{\ec}{\end{center}}
\newcommand{\bn}{\begin{enumerate}}
\newcommand{\en}{\end{enumerate}}
\newcommand{\bi}{\begin{itemize}}
\newcommand{\ei}{\end{itemize}}
\newcommand{\bqr}{\begin{eqnarray}}
\newcommand{\eqr}{\end{eqnarray}}
\newcommand{\bfig}{\begin{figure}[tbp]}
\newcommand{\efig}{\end{figure}}
\newcommand{\btab}{\begin{table}[ht]}
\newcommand{\etab}{\end{tabular}\ec\end{table}}
\newcommand{\bl}{\begin{large}}
\newcommand{\el}{\end{large}}

\newcommand{\nb}{\nonumber}

\newcommand{\Ps}{P_{\sigma}}

\newcommand{\de}{\delta}

\newcommand{\vr}{\hbox{\bf r}}

\newcommand{\nuc}[2]{\mbox{\relax\ifmmode{}^{#1}{\protect\text{#2}}\else${}^{#1}$#2\fi}}
\title      {Pairing correlations. \\
	    Part 2: Microscopic analysis of odd-even mass staggering in nuclei.}
\author     {
             T. Duguet, P. Bonche, \\
             {\em Service de Physique Th\'eorique, CEA Saclay,} \\
             {\em 91191 Gif sur Yvette Cedex, France} \\
             P.-H. Heenen, \\
	     {\em Service de Physique Nucl\'eaire Th\'eorique, Universit\'e Libre} \\ 
	     {\em de Bruxelles, C.P 229, B-1050 Bruxelles, Belgium} \\
\and	     J. Meyer \\
             {\em Institut de Physique Nucl\'eaire de Lyon, } \\
	     {\em CNRS-IN2P3 / Universit\'e Claude Bernard Lyon 1,} \\
	     {\em 43, Bd. du 11.11.18, 69622 Villeurbanne Cedex, France}
            }% end author

\begin{document}
 
\maketitle
 
\begin{abstract}

The odd-even mass staggering in nuclei is analyzed in the context of 
self-consistent mean-field calculations, for spherical as well as 
for deformed nuclei. For these nuclei, the respective merits of
the energy differences $\Delta^{(3)}$ and 
$\Delta^{(5)}$  to extract both the 
pairing gap and the time-reversal symmetry breaking effect at the same time
are extensively discussed. 
The usual mass formula $\Delta^{(3)}$, is shown to contain additional 
mean-field contributions when realistic pairing is used in the calculation. 
A simple tool is proposed in order to remove time-reversal symmetry 
breaking effects from $\Delta^{(5)}$. 
Extended comparisons with the odd-even mass staggering obtained 
in the zero pairing limit (schematic model and self-consistent calculations) 
show the non-perturbative contribution of pairing correlations on this 
observable.



\end{abstract}

\footnote{Corresponding author~: duguet@spht.saclay.cea.fr}

%\newpage

\section{Introduction}
\label{intro}

The Odd-Even Staggering (OES) of binding energies is a common phenomenon of several finite many-fermion systems. In nuclei, it has been attributed to an experimental evidence of pairing correlations\cite{Boh8}. Assuming that masses are smooth functions of the number of neutrons and protons except for pairing effects, simple expressions have been derived for the gap parameter $\Delta$ based on binding energy differences between even and odd neighboring nuclei\cite{BM75,jensen}. Detailed analyses\cite{mad} and pairing adjustments\cite{mol} have been based on these expressions. The simplest example is the well-known three-point mass formula: 

\begin{equation}
\Delta^{(3)}(N) = \frac{(-1)^{N}}{2} \, [E(N\!+\!1)-2E(N)+E(N\!-\!1)] \, \, \, , \\  \vspace{0.3cm}
\label{defd3}
\end{equation}
where $N$ is the number of nucleons (neutrons or protons).

A study of the OES in light alkali-metal clusters and in light $N$ = $Z$ nuclei~\cite{Haa98} has led to the conclusion that this phenomenon was not due to pairing correlations but rather to deformation effects (Jahn-Teller OES)~\cite{cle,man}. That work motivated a study by Satula {\it et al.}~\cite{SDN98} on the mean-field contribution to the OES in nuclei, especially coming from deformation. To isolate mean-field effects, the pairing force was set to zero and Hartree-Fock (HF) calculations were performed for light deformed nuclei. In this context, they observed an OES of the energy through $\Delta^{(3)}$, itself oscillating: 

\begin{eqnarray}
&\begin{array}{lccc}
\Delta^{(3)}_{HF}(N) &\approx& 0 & \, {\rm if} \ N \ {\rm is \ odd} \, \,  , \\
 & \approx & \frac{e_{k}-e_{k-1}}{2} & \, {\rm if} \ N \ {\rm is \ even}  \, \,  .
\end{array}&
\label{delta3hf}
\end{eqnarray}

\begin{figure}
\begin{center}
\leavevmode
\centerline{ \psfig{figure=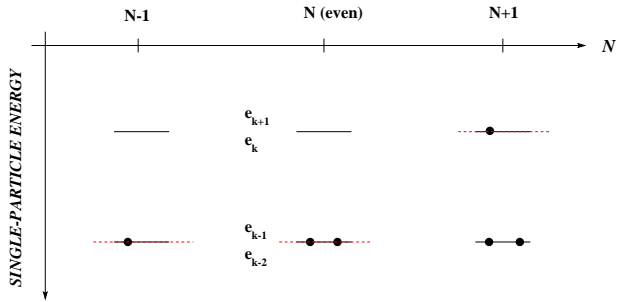,height=4cm}}
\end{center}
\caption{Schematic single-particle spectra and occupations for three successive deformed nuclei.}
\label{spectre}
\end{figure}

For even $N$, $(e_{k}-e_{k-1})$ is the gap around the Fermi level in the single-particle spectrum. It is zero for spherical nuclei (apart across sub-shells) and differs from zero for deformed nuclei because of the spread doubly-degenerate spectrum (see Fig.~\ref{spectre}). Deformation is thus found to be responsible for a direct contribution to the three point odd-even mass formula.  

With pairing correlations, $\Delta^{(3)}(odd)$ can a priori be a measure of pairing effects only,  whereas $\Delta^{(3)}(even)$ contains an additional contribution related to the splitting of the single-particle spectrum around the Fermi level. Such a scheme cannot account for the same oscillation of $\Delta^{(3)}$ in spherical nuclei because of the large degeneracy in spherical shells.

On the other hand, recent calculations of spherical tin isotopes including pairing (HF+BCS calculations)~\cite{bend} have led to the conclusion that the five-points formula $\Delta^{(5)}$ was in this case a better approximation of the pairing gap that $\Delta^{(3)}(odd)$.

The purpose of the present study is to analyze, for spherical and deformed nuclei, the different contributions to odd-even mass differences in a fully self-consistent mean-field picture including time-odd components of the force. Our aim is to give a coherent picture and to extract a quantity more directly related to pairing correlations. In addition, we want to investigate the connection between results obtained with and without the inclusion of pairing (cf. Eq.~\ref{delta3hf}), as well as the physical content of this connection.

The present work is based on the conclusions of Ref.~\cite{dug3} (hereafter refereed as paper I), and is organized as follow. In section~\ref{secfull} we introduce a separation of the microscopic binding energy which allows to separate different types of contributions to odd-even mass formulas. The theoretical framework used to perform our mean-field calculations is detailed in section~\ref{subsecframe}. Results on spherical as well as deformed nuclei are presented in section~\ref{secres}. The evolution of odd-even mass differences as a function of pairing correlations intensity is studied in section~\ref{transition} using a schematic BCS model and self-consistent calculations. Finally, the analysis of the results and the conclusions are given in section~\ref{secconclu}.

\section{Odd-even mass differences in self-consis\-tent mean-field calculations}
\label{secfull}

To evaluate and understand the contributions to the odd-even mass differences of nuclei in a fully self-consistent mean-field picture, two questions are addressed in what follows:
\bi
\item[(1)] how to define a procedure to extract different contributions to the OES and to identify unambiguously their physical content? \\
\item[(2)] is the analysis of the odd-even mass differences at the HF level of any help to understand what happens in presence of pairing correlations? 
%In other words, it is essential to understand whether the analysis of the OES at the HF level is physically %grounded as regards the typical value of the pairing gap in atomic nuclei.
\ei

\subsection{Smooth contribution to mass formulas}
\label{subsecoddeven}

Several finite-difference mass formulas \cite{jensen,mad,mol} are used to evaluate the neutron or proton ``pairing gaps''. The aim is to extract the quickly varying part of the energy as a function of some parameters such as the number of neutrons or protons. The underlying assumption is that the microscopic energy splits into a quickly varying part and a smooth one. In the description of nuclear structure, the rapidly varying component of the energy can be related to different phenomena such as shell closures, $N$ = $Z$ line, light mass nuclei, time-reversal symmetry breaking and reduction of pairing by blocking in odd nuclei. The OES being related to the last two effects, appropriate mass regions must be chosen in order to avoid the first three ones.

We define the smooth part of the energy as the one obtained when all nuclei are calculated as if they were even ones (no blocking and no breaking of time-reversal invariance in odd nuclei). Such an energy should not undergo odd-even irregularities. It will be referred to as E$^{HFBE}$ for ``Hartree-Fock-Bogolyubov {\it Even}'' and the associated wave-function will be denoted ${\rm \mid \Psi^{HFBE} >}$. Such a definition of the smooth part of the microscopic binding energy has already been used in a work dealing with the OES in odd nuclei~\cite{bend}. Then, the energy of an odd nucleus can be written as: 

\begin{eqnarray}
E^{HFB}(N) &=& E^{HFBE}(N) + [E^{HFB}(N) - E^{HFBE}(N)] \nb  \\
&=& E^{HFBE}(N) + \, \, \, \,  \overbrace{E^{pol}(N) \, \, \, \, \, + \, \, \, \, \, \, \, \Delta(N)\,}  \label{defener} \, \, \, \, \, \, ,
\end{eqnarray}
where $E^{pol}(N)$ is the difference of binding energy due to the time-reversal symmetry breaking effect, and $\Delta(N)$ is the positive contribution related to the fact that the odd nucleon is unpaired in the final HFB one quasi-particle (qp) state. It is denoted as the self-consistent pairing gap whereas $E^{pol}(N) + \Delta(N)$ is the self-consistent qp energy~\cite{dug3}.

We shown in Ref.~\cite{dug3} that the state ${\rm \mid \Psi^{HFBE} >}$ arises naturally as an intermediate step in the nucleon addition process. It defines an underlying even structure in an odd nucleus.

\subsection{Finite difference mass formulas}
\label{subsecformulas}

Starting from the separation procedure defined by Eq.~\ref{defener}, the smooth part of the binding energy can be expanded in a power serie around a given mass number $N_0$:

\be
E^{HFB}(N) = \sum_{k=0}^{\infty} \, \frac{1}{k!} \, \left. \frac{\partial^{k} E^{HFBE}}{\partial N^{k}} \, \right|_{N_0} \, (N - N_0)^{k} \, + \, E^{pol}(N) \, + \, \Delta(N)  \vspace{0.3cm}
\label{serie}
\ee

Finite-difference formulas have been derived \cite{jensen,mad,mol} to eliminate the successive derivatives of the smooth part of the energy. The three-point difference is written as:

\begin{eqnarray}
\Delta^{(3)}_{HFB}(N_0) &=&  \Delta^{(3)}_{HFBE}(N_0) + \Delta^{(3)}_{pairing+pol}(N_0) \, \, .
\label{d3}
\end{eqnarray}

\noindent Using Eq.~\ref{serie}, we have:

\be
\Delta^{(3)}_{HFBE}(N_0) \approx  \frac{(-1)^{N_0}}{2} \, \left. \frac{\partial^{2} E^{HFBE}}{\partial N^{2}} \, \right|_{N_0} \,  \,  ,
\label{d3approx}
\ee
and
\vspace{0.5cm}

\begin{eqnarray}
&\begin{array}{rclcl}
\Delta^{(3)}_{pairing+pol}(N_0) \!&=& \! \Delta(N_0)+E^{pol}(N_0)  &\qquad&{\rm if} \ N_0 \ {\rm is \ odd,}  \\ [8pt]  
\!&=& \! \frac{1}{2} \left\{ \Delta(N_0\!-\!1)+\Delta(N_0\!+\!1) \right. && \\
\!&& \, \, \, \, \left. +E^{pol}(N_0\!-\!1)+E^{pol}(N_0\!+\!1) \right\} & &{\rm if} \ N_0 \ {\rm is \ even.} 
\end{array}&  \nb 
\end{eqnarray}
\vspace{0.5cm}

Similar expressions are obtained for the fourth-order formula (five points difference):

\begin{eqnarray}
\Delta^{(5)}_{HFB}(N_0) &=& - \frac{(-1)^{N_0}}{8} \, [ E^{HFB}(N_0\!+\!2) - 4E^{HFB}(N_0\!+\!1) \nb \\
&& + \, \, 6E^{HFB}(N_0) - 4E^{HFB}(N_0\!-\!1) + E^{HFB}(N_0\!-\!2) ] \, \\
&=& \Delta^{(5)}_{HFBE}(N_0) + \Delta^{(5)}_{pairing+pol}(N_0) \,  \,  \,  \, . \nb
\label{d5}
\end{eqnarray}
\vspace{0.4cm}

Higher order formulas can be derived in the same way.

\subsection{Remarks on $E^{HFBE}$ and $\Delta(N)+E^{pol}(N)$}
\label{subsecexpect}

If $E^{HFBE}$ represents a smooth part of the binding energy, 
its contribution to $\Delta^{(n)}(N)$ 
should be of the same order of magnitude (at least in absolute value) 
for odd and even neighbor nuclei. 
For instance, the contribution to $\Delta^{(3)}_{HFB}(N)$ is approximately 
equal to $(-1)^{N}/2 \, \partial^{2} E^{HFBE}/\partial N^{2} \, \,$ ,
and since this second derivative is always positive and smooth, 
it  gives an oscillating contribution of similar amplitude for 
odd and even $N$. 

Moreover, if such an hypothesis is valid, the contribution of $E^{HFBE}$ to $\Delta^{(n)}(N)$ will tend to zero, and $\Delta^{(n)}_{HFB}(N)$ to $\Delta^{(n)}_{pairing+pol}(N) \,$, 
with increasing order $\, n$ of the finite difference formula. The decrease of $\Delta^{(n)}_{HFBE}$ with $n$ will have to be checked in order to validate the above energy separation. 

With increasing $n$, $\Delta^{(n)}_{pairing+pol}(N)$ is an average of $\Delta + E^{pol}$ over a larger number of nuclei. 
This assumes that $\Delta + E^{pol}$ varies slowly with the (odd) nucleon number. If so, $\Delta^{(n)}_{pairing+pol}(N)$ will be a relevant observable in the nucleus with $N$ nucleons. 

The choice of the order $n$ of the formula used to extract a ``pairing gap'' $\Delta(N)$ through the odd-even staggering should be a compromise as regards to the last two remarks.

\section{Theoretical framework}
\label{subsecframe}

In order to study the odd-even mass staggering in the framework of self-consistent calculations, one has to use the Hartree-Fock-Bogolyubov (HFB) 
approximation, since time-reversal symmetry is lost in odd nuclei 
because of the blocking of a single nucleon.
This implies that time-odd components of the effective interaction
must be included in the calculations. 
They are not as well determined~\cite{doba2} 
as the even components 
since nothing constrains them specifically 
in the usual fitting procedure of effective forces~\cite{chab}. 
Despite these uncertainties, it is important 
to study their contributions to binding energies and odd-even mass 
differences which are known to be significant \cite{rutz,xu}.

The general formalism used here is detailed in Ref.~\cite{tera}. It is based on the  self-consistent Hartree-Fock Bogolyubov method, with an approximate particle number projection by the Lipkin-Nogami prescription. In the particle-hole channel, we use a two-body force of the Skyrme type, SLy4, which has been adjusted to reproduce also the characteristics of the infinite neutron matter and, consequently, should have good isospin properties~\cite{chab}.  This force has been shown to describe satisfactorily nuclear properties for which it had not been adjusted such as  super-deformed rotational bands \cite{rigol,fallon} and the structure and decay of super-heavy elements \cite{cwiok1}. It should be mentioned that the time-odd components of the force are deeply involved in the description of rotational properties. The capacity of SLy4 to reproduce these observables is an advantage as regards to the previous discussion. In the $T$ = 1 particle-particle channel, we use a surface-peaked delta force (Eq. \ref{potsurf}) adjusted on the low spin behavior of the moments of inertia of super-deformed bands in the A $\approx$ 150 region \cite{rigol}. This pairing force has also been shown to work well in very different mass regions, up to the transfermium one, to describe ground state as well as rotational properties \cite{dug}. It is given by

\be
\hat{V_{\tau}} =  \frac{V_{\tau}}{2} \, \, (1 - \Ps) \, \, \de (\vr_1 - \vr_2 )  \, \, (1 - \frac{\rho(\vec R)}{\rho_c}) \, , \vspace{0.3cm}
\label{potsurf}
\ee
where $V_{\tau}$ = $-$ 1250 MeV.fm$^{-3}$ ($\tau$ stands for proton or neutron), $\rho(\vec R)$ is the local matter density of the nucleus, $\Ps$ is the spin exchange operator and $\rho_c$ = 0.16 fm$^{-3}$ the nuclear saturation density. As $\hat{V_{\tau}}$ is a contact interaction, we use a cut-off for the active pairing space which ranges from 5 MeV below to 5 MeV above the Fermi level~\cite{bonche} in the single-particle spectrum.

\section{Results}
\label{secres}

\subsection{Spherical nuclei}
\label{subsecressphere}

Seventy ground-states have been calculated along the tin isotopic chain, from $^{100}$Sn to $^{169}$Sn. Each odd $N$ nucleus has been calculated twice: first, 
as a HFB fully paired vacuum with an odd average number of neutrons (HFBE state), then with the fully self-consistent HFBLN scheme. In this case, several one qp configurations are investigated  
to determine the configuration corresponding to the ground state. Due to the magic number of proton ($Z$ = 50), all these nuclei are found to be spherical in our three-dimensional lattice calculation.

\begin{figure}
\begin{center}
\leavevmode
\centerline{\psfig{figure=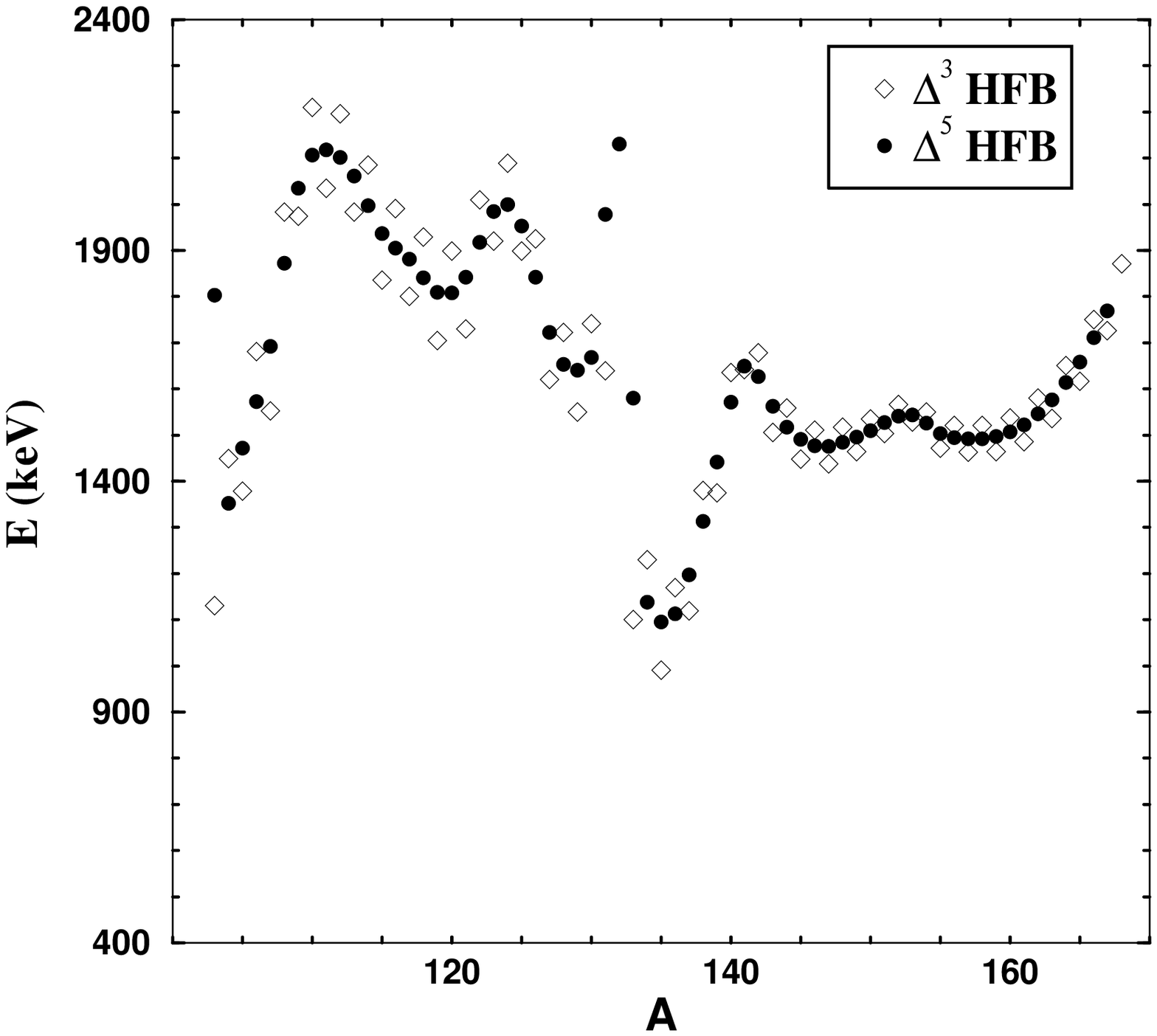,height=6cm} \psfig{figure=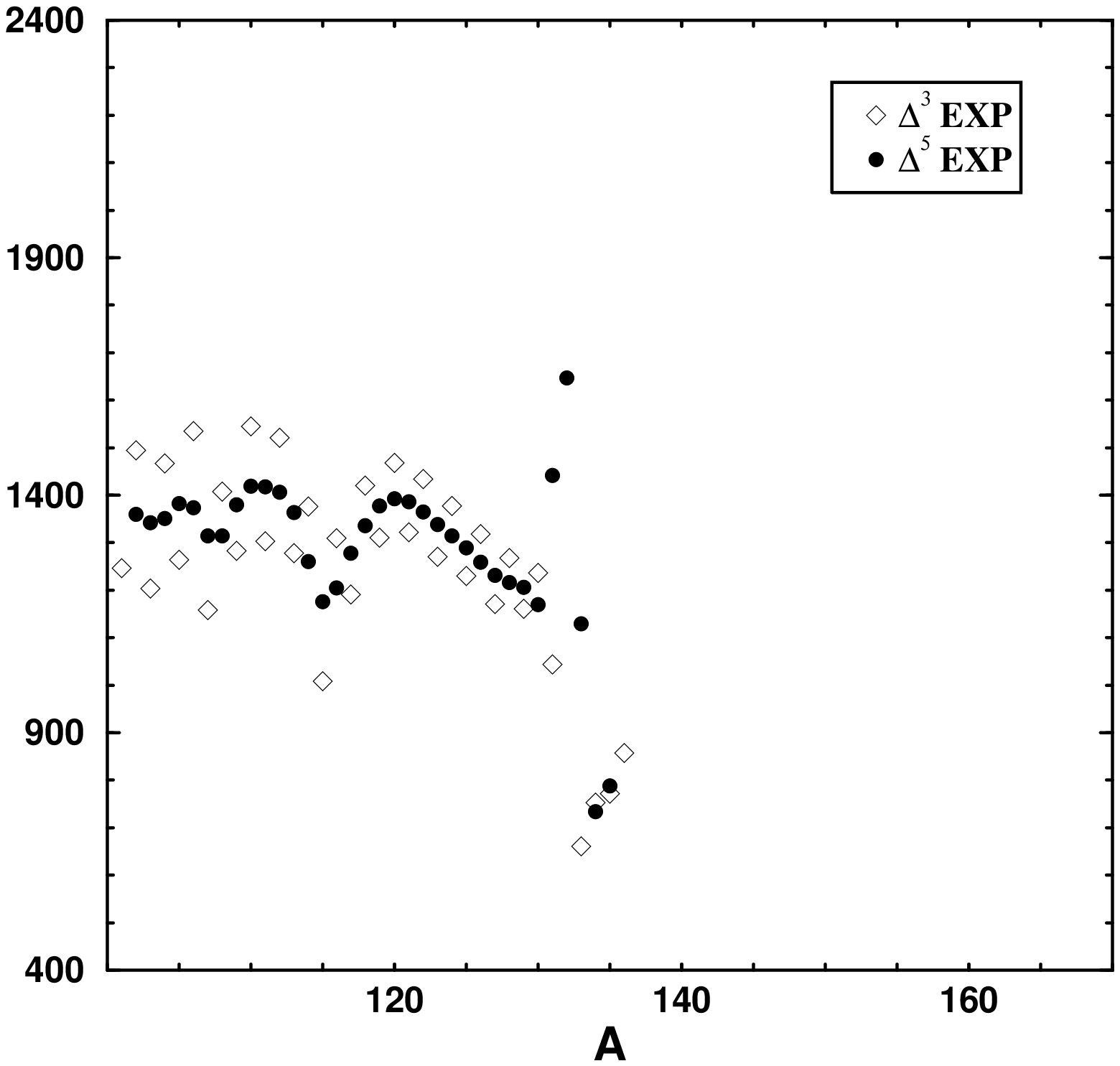,height=6cm}}
\end{center}
\caption{Left: calculated odd-even mass differences $\Delta^{(3)}_{HFB}$ and $\Delta^{(5)}_{HFB}$ for the tin isotopic line from $^{100}$Sn to $^{169}$Sn. Right: known $\Delta^{(3)}_{Exp}$ and $\Delta^{(5)}_{Exp}$. Experimental data are taken from~\cite{audi}.}
\label{deltashfbSn}
\end{figure}

Calculated and experimental~\cite{audi} $\Delta^{(3)}(N)$ and $\Delta^{(5)}(N)$ along this chain are given on Fig.~\ref{deltashfbSn}. The staggering of $\Delta^{(3)}(N)$ can be seen on the two panels, the values for odd $N$ being smaller than for  even $N$. The amplitude of this staggering is smaller for neutron-rich nuclei. No staggering occurs for $\Delta^{(5)}(N)$. 

Calculated $\Delta^{(3,5)}(N)$ are greater than experimental ones by several hundreds keV while the fit used for the pairing force has been shown to work well in reproducing rotational properties in several mass regions. This is partly due to numerical difficulties to converge very well several odd nuclei 
in the region $^{107-125}$Sn which leads to an underestimation of their binding
energy by two to three hundreds keV. Moreover, correlations 
beyond mean-field in odd nuclei  should be larger than in even ones 
because of the large number of low-lying individual excited states 
in the first case. Treating explicitly the residual interaction through configuration mixing in odd and even nuclei is expected to  lower the OES by approximately three hundreds keV~\cite{dech}.

\begin{figure}
\begin{center}
\leavevmode
\centerline{\psfig{figure=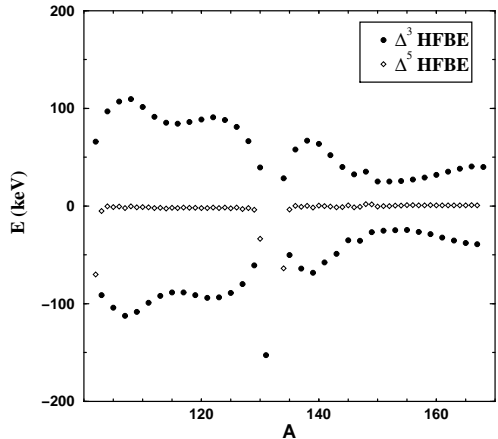,height=6cm} }
\end{center}
\caption{Calculated odd-even mass differences $\Delta^{(3)}_{HFBE}$ and $\Delta^{(5)}_{HFBE}$ for the tin isotopic line from $^{100}$Sn to $^{169}$Sn.}
\label{deltashfbpSn}
\end{figure}

On Fig.~\ref{deltashfbpSn} are plotted the contributions 
$\Delta^{(3)}_{HFBE}$ and $\Delta^{(5)}_{HFBE}$ to $\Delta^{(3)}_{HFB}$ 
and $\Delta^{(5)}_{HFB}$ respectively. Apart for the magic number $N=82$, one has
$|\Delta^{(3)}_{HFBE}| \gg  |\Delta^{(5)}_{HFBE}|=0$. 
This justifies the identification of E$^{HFBE}$ as the smooth part 
of the energy, and indicates that higher order 
formulas are not needed.

Assuming that $\Delta^{(3)}_{pairing+pol}$ is equal to
$\Delta^{(5)}_{pairing+pol}$ since $\Delta(N) + E^{pol}(N)$ is constant over a few odd nuclei (see section \ref{subsecoddeven}), one can write:

\begin{eqnarray}
\Delta^{(5)}_{HFB}(N) &\approx& \Delta^{(5)}_{pairing+pol}(N)  \nb  \\
\Delta^{(3)}_{HFB}(N) -  \Delta^{(5)}_{HFB}(N) &\approx& \Delta^{(3)}_{HFBE}(N)  \label{staggd3}  \, .
\end{eqnarray}

\begin{figure}
\begin{center}
\leavevmode
\centerline{ \psfig{figure=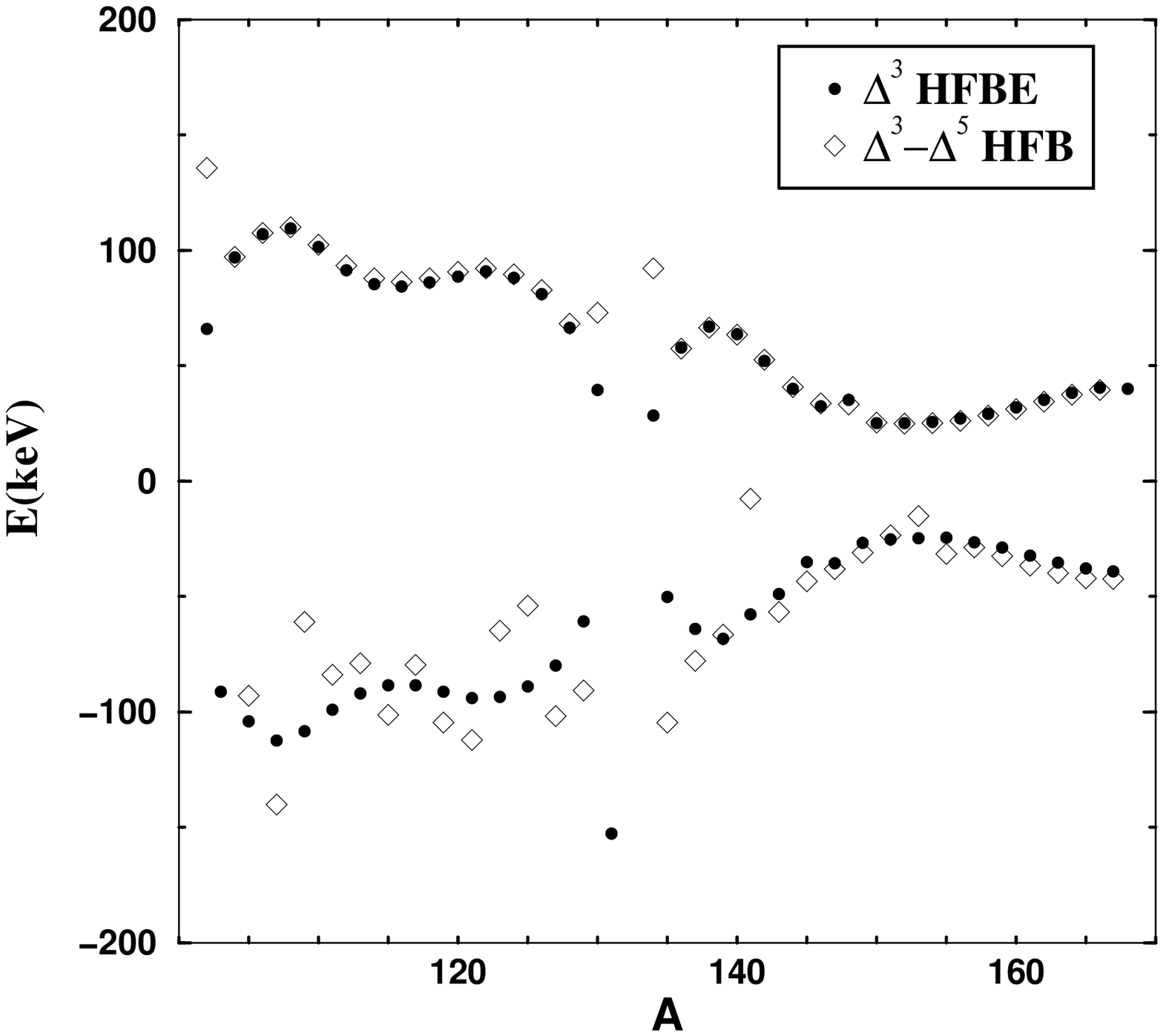,height=6cm} \psfig{figure=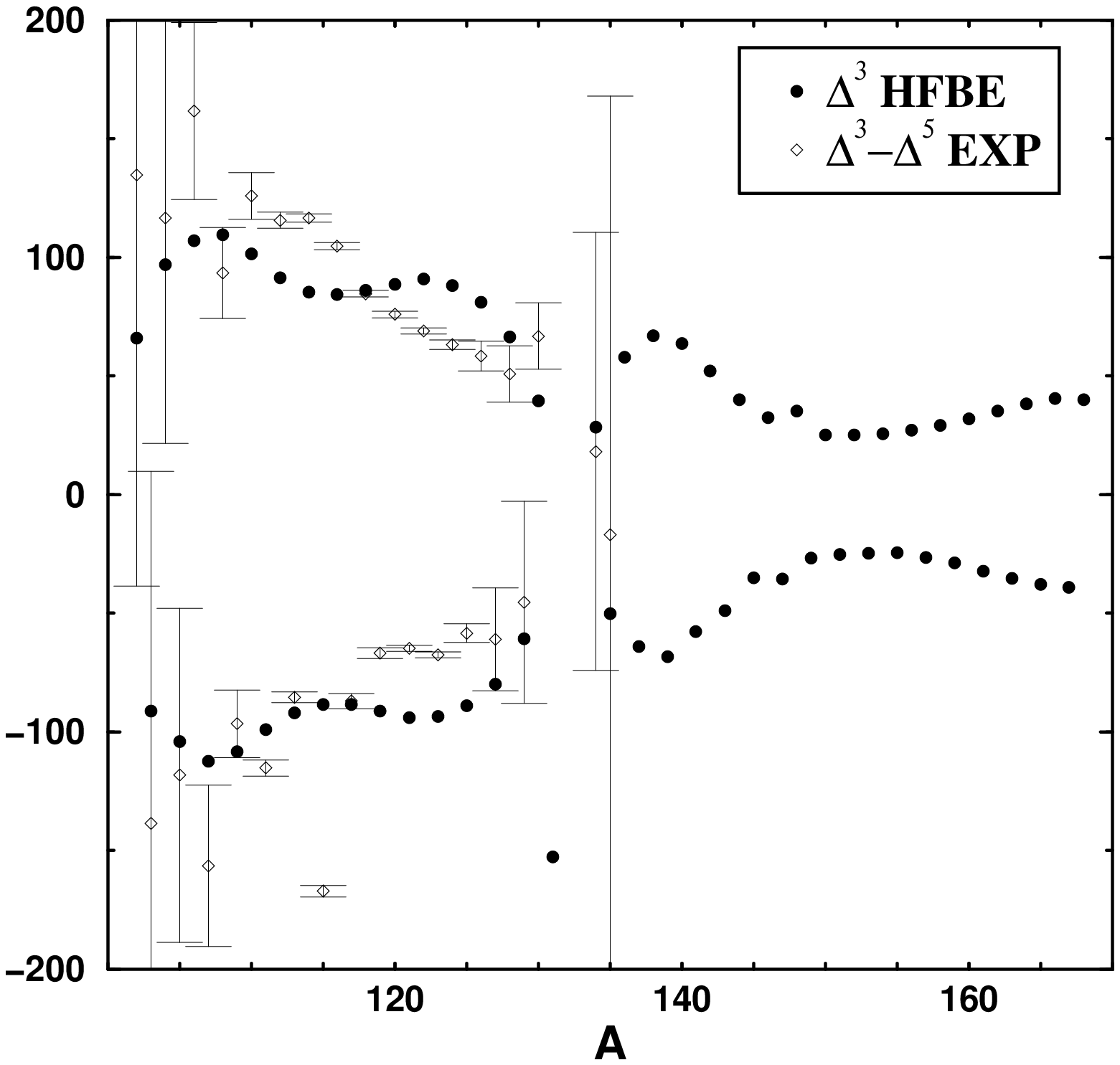,height=6cm}}
\end{center}
\caption{Left: $\Delta^{(3)}_{HFBE}$ is compared to $\Delta^{(3)}_{HFB}$ - $\Delta^{(5)}_{HFB}$ along the tin isotopic line. Right: comparison of $\Delta^{(3)}_{HFBE}$ with experiment (see text).}
\label{d3-d5hfbSn}
\end{figure}

The two sides of Eq. \ref{staggd3} are compared on the left handside of Fig.~\ref{d3-d5hfbSn}. The accuracy of the agreement is impressive along the whole isotopic line (except as expected for the magic number $N$ = 82). 
It confirms that the contribution of the smooth part 
of the energy to $\Delta^{(3)}$ is equal in absolute value 
for even and odd $N$ (see section \ref{subsecoddeven}),
the sign $(-1)^{N}$ being responsible for the oscillating pattern 
of this contribution. The $\Delta^{(3)}$ staggering 
corresponds to an oscillation around $\Delta^{(5)}$ due 
to $\Delta^{(3)}_{HFBE}$. One can therefore conclude that 
$\Delta^{(5)}$ is a  measure of the rapidly varying part 
of the energy $\Delta^{(5)}_{pairing+pol}(N) \approx \Delta(N) + E^{pol}(N)$. 
On the contrary, $\Delta^{(3)}(odd)$ 
contains smooth contributions, in particular 
the full asymmetry energy contribution to the OES.

The right panel of Fig.~\ref{d3-d5hfbSn} provides a comparison between 
$\Delta^{(3)}_{HFBE}$ and $\Delta^{(3)}_{Exp}$ - $\Delta^{(5)}_{Exp}$. 
The agreement is very good. 
Although the absolute values of $\Delta^{(3,5)}_{Exp}$
are overestimated by a few hundred keV, 
the staggering of $\Delta^{(3)}_{Exp}$ around $\Delta^{(5)}_{Exp}$ is 
well reproduced by the contribution coming from the smooth part 
of the energy $\Delta^{(3)}_{HFBE}$. This 
shows the strong decoupling between the two contributions to Eq.~\ref{d3}.

\begin{figure}
\begin{center}
\leavevmode
\centerline{\psfig{figure=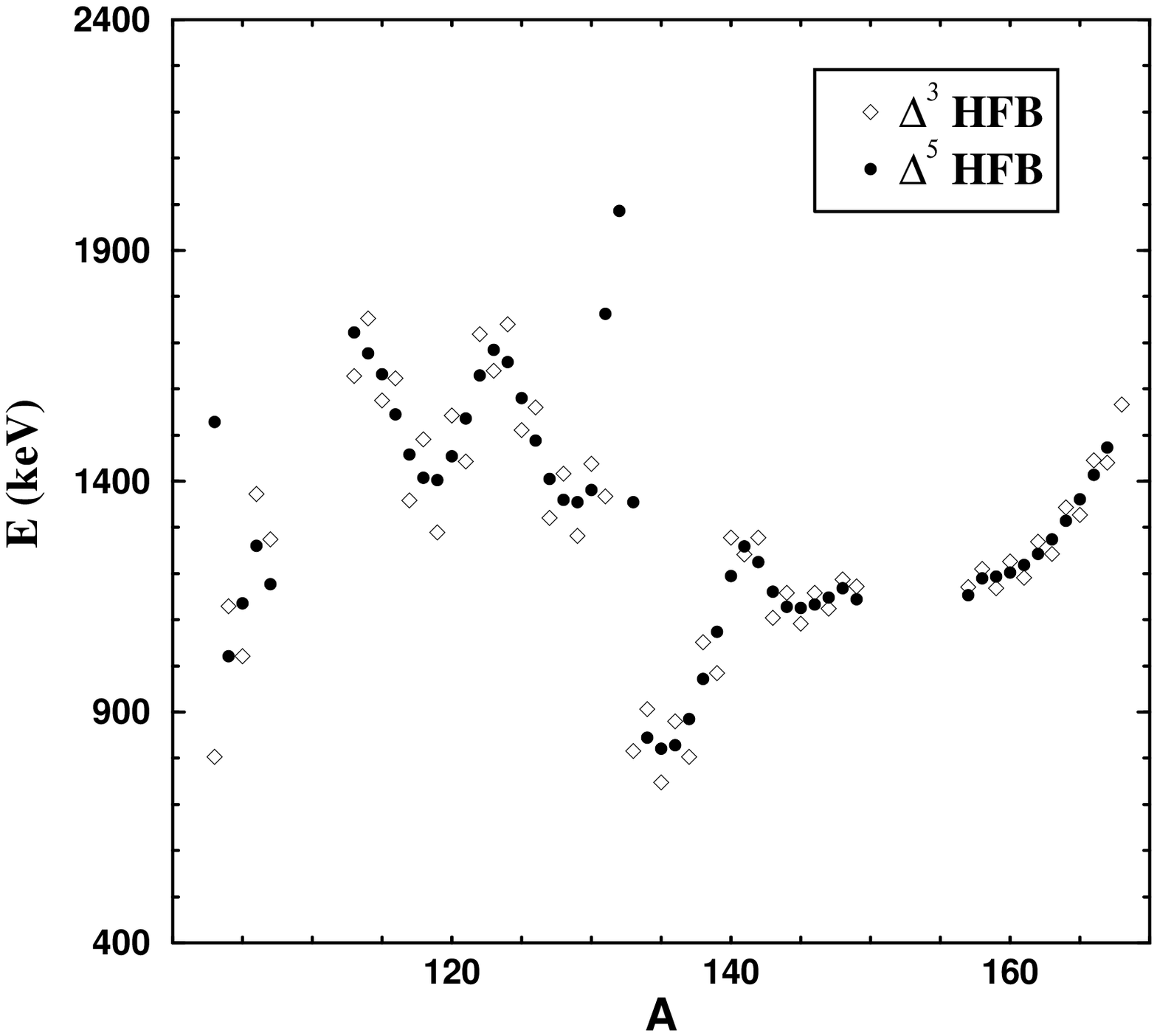,height=6cm} \psfig{figure=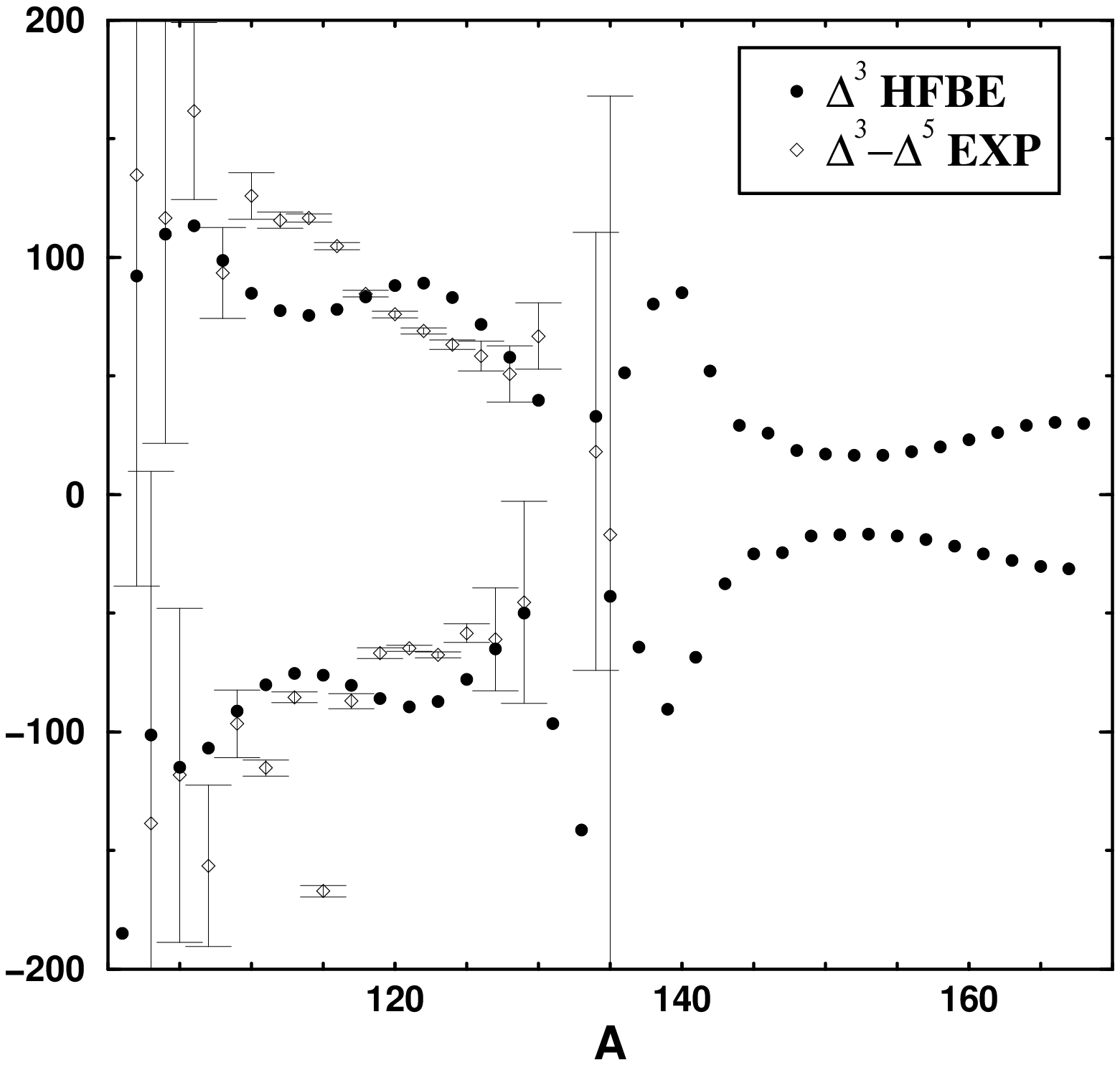,height=6cm}}
\end{center}
\caption{Left: calculated odd-even mass differences $\Delta^{(3)}_{HFB}$ 
and $\Delta^{(5)}_{HFB}$ for the tin isotopic line 
with a reduced neutron pairing strength $V_{n}$ = $-$ 1000 MeV.fm$^{-3}$. 
Right: $\Delta^{(3)}_{HFBE}$ is compared to $\Delta^{(3)}_{EXP}$ - 
$\Delta^{(5)}_{EXP}$ for the same $V_{n}$.}
\label{deltashfbSn3}
\end{figure}

To confirm this decoupling, we have performed
the same calculation with a  decrease of the neutron pairing intensity 
$V_{n}$ by  20 $\%$ to $ - 1000$~MeV.fm$^{-3}$. 
The left panel of Fig.~\ref{deltashfbSn3} shows the response of the absolute 
OES $\Delta^{(3,5)}_{HFB}$. A few points are missing on the figure due to problems of convergence for
some nuclei with a reduced pairing strength. The odd-even mass differences $\Delta^{(3)}_{HFB}$ and $\Delta^{(5)}_{HFB}$ which include the contribution coming from the pairing gap, decrease also
by approximately 20 $\%$. On the other hand, the right panel of Fig.~\ref{deltashfbSn3} shows that 
the oscillation of $\Delta^{(3)}_{HFB}$ due to the smooth part of the energy 
is not modified. Thus, the separation of Eq.~\ref{defener} divides the energy 
in a part extracted through $\Delta^{(5)}$ responding directly to the pairing intensity and in another one coming from E$^{HFBE}$ (almost) insensitive to it. 

In a previous study of $^{254}$No~\cite{dug},
we have shown that the dynamical moment of inertia of its ground state rotational band
depends more strongly on the radial dependence of the pairing 
force than on its intensity for a given radial form factor. 
On the contrary, the OES around $^{254}$No has been found 
to vary proportionally to a change of the intensity of the pairing. 
From this and from the present results, we can conclude that
the part of the pairing energy contained in the even energy E$^{HFBE}$ is probed by 
observables involving the nucleus as a whole, such as rotation, and 
is more sensitive to the analytical structure of the force.
On the other hand, the part defined by $\Delta(N)$ is 
related to a specific blocked orbit and is very sensitive to the 
intensity of the pairing force. This provides with two different observables
to adjust the strength and the radial dependence of the pairing force.

\subsection{Deformed nuclei}
\label{subsecresdef}

\begin{figure}
\begin{center}
\leavevmode
\centerline{\psfig{figure=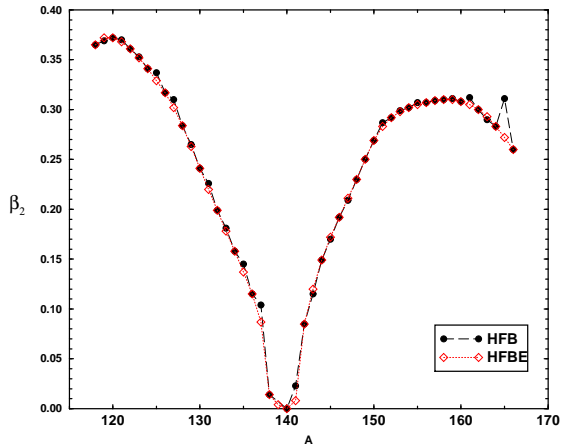,height=6cm}}
\end{center}
\caption{Quadrupole axial deformation as a function of the mass number along the cerium isotopic chain for HFB and HFBE calculations.}
\label{beta2}
\end{figure}

Let us now extend our analysis to deformed nuclei. Forty nine Cerium isotopes 
have been calculated, from $^{118}$Ce to $^{166}$Ce. The deformation parameter,
 $\beta_{2}$, is given as a function of $A$ on Fig.~\ref{beta2} for HFB and 
HFBE calculations. The ground-state quadrupole deformation 
undergoes large variations from a region of strong prolate 
deformation around  $^{118}$Ce ($\beta_{2} \approx 0.37$) to a prolate 
deformation region ($\beta_{2} \approx 0.31$ around $^{160}$Ce) through 
the spherical $^{140}$Ce nucleus. 
Fig.~\ref{beta2} illustrates the fact that $\beta_{2}^{HFBE}$ reproduces 
the mean evolution of $\beta_{2}^{HFB}$ with $A$.

It has been shown in paper I that, apart for the time-reversal 
symmetry breaking effect, the HFBE and HFB calculations lead to nearly
identical mean-fields. The influence of the deformation 
on the binding energy of odd nuclei is contained in E$^{HFBE}$ and the 
Jahn Teller contribution to the odd-even mass differences 
$\Delta^{(n)}_{HFB}$ will thus be extracted through $\Delta^{(n)}_{HFBE}$. 
The energy separation~\ref{defener} should allow
to isolate the self-consistent qp energy through odd-even mass formulas 
in the same way as for spherical nuclei. 
Let us illustrate these statements along the cerium isotopic line.

\begin{figure}
\begin{center}
\leavevmode
\centerline{\psfig{figure=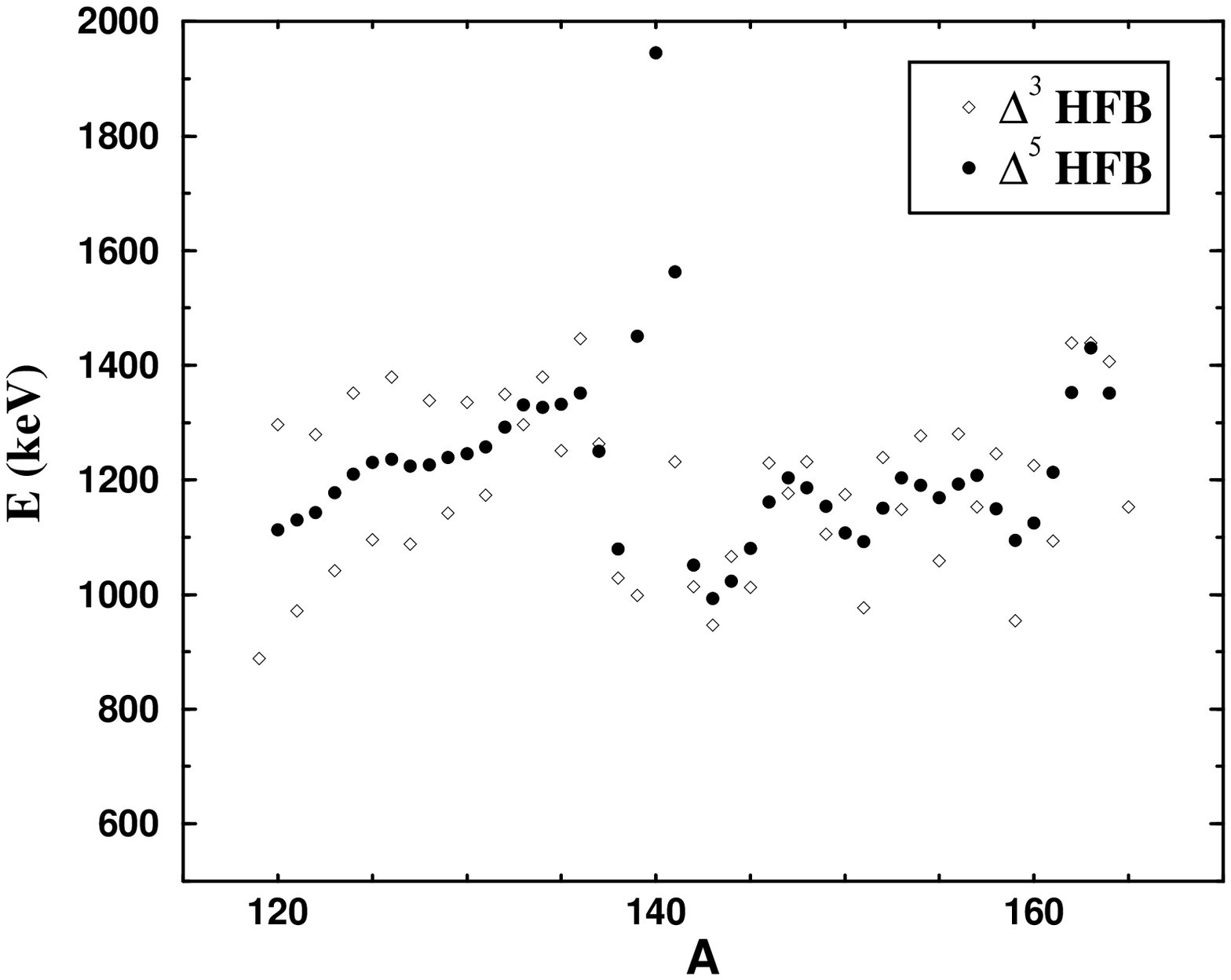,height=6cm} \psfig{figure=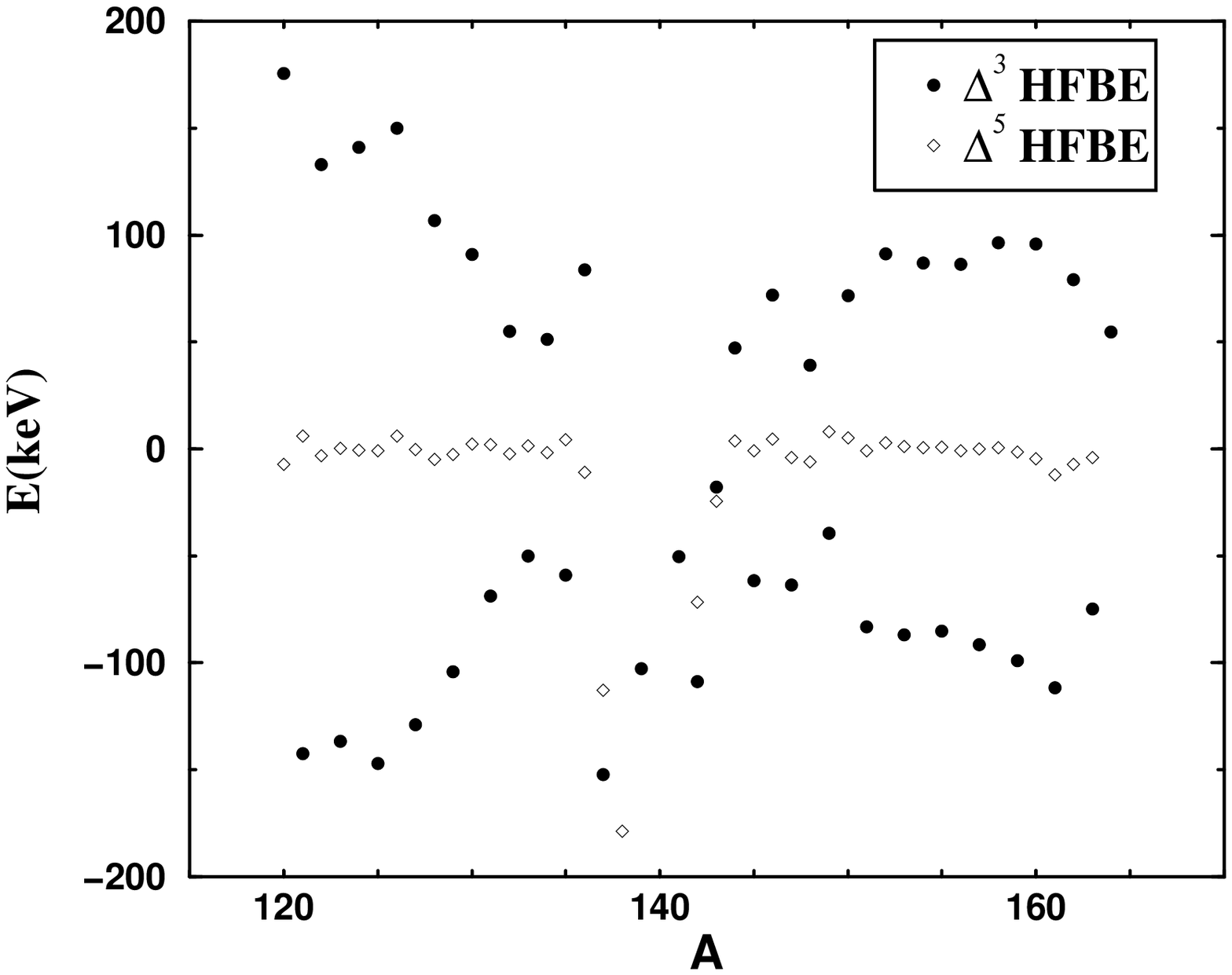,height=6cm}}
\end{center}
\caption{Left: calculated odd-even mass differences $\Delta^{(3)}_{HFB}$ 
and $\Delta^{(5)}_{HFB}$ for the cerium isotopic line 
from $^{118}$Ce to $^{166}$Ce. 
Right: calculated odd-even mass differences 
$\Delta^{(3)}_{HFBE}$ and $\Delta^{(5)}_{HFBE}$.}
\label{deltashfbCe}
\end{figure}

\begin{figure}
\begin{center}
\leavevmode
\centerline{ \psfig{figure=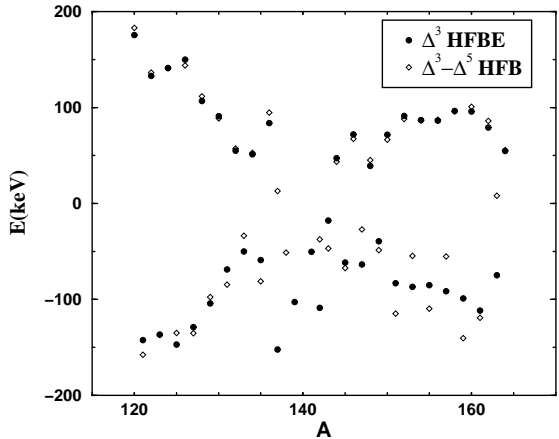,height=6cm} }
\end{center}
\caption{Comparison between calculated $\Delta^{(3)}_{HFBE}(N)$ (filled circles) and $\Delta^{(3)}_{HFB}(N) -  \Delta^{(5)}_{HFB}(N)$ (empty diamonds) along the cerium isotopic line.}
\label{d3-d5hfbCe}
\end{figure}

On the left (right) panel of Fig.~\ref{deltashfbCe} are plotted 
$\Delta^{(3)}_{HFB(E)}$ 
and $\Delta^{(5)}_{HFB(E)}$. 
The comparison between $\Delta^{(3)}_{HFB}-\Delta^{(5)}_{HFB}$ 
and $\Delta^{(3)}_{HFBE}$ is presented on Fig.~\ref{d3-d5hfbCe}. 
Similar results and agreements as for tin isotopes are 
obtained: $|\Delta^{(3)}_{HFBE}|$ is much larger than  $|\Delta^{(5)}_{HFBE}|$
which is close to zero  
and $\Delta^{(3)}_{HFB}(N)-\Delta^{(5)}_{HFB}(N)$ is approximately
equal to $\Delta^{(3)}_{HFBE}(N)$. 
The contribution $\Delta(N) + E^{pol}(N)$ to the odd HFB energy 
is  extracted through $\Delta^{(5)}$ 
for deformed nuclei as for spherical ones. 
The effect of deformation on the OES (Jahn Teller OES) is 
included in the contribution coming from E$^{HFBE}$ which 
reproduces the oscillation 
of $\Delta^{(3)}$ around $\Delta^{(5)}$ (Fig.~\ref{d3-d5hfbCe}). 
This contribution is identical for odd and 
even neighboring nuclei, with an opposite sign contrary to what was found in
Ref.~\cite{SDN98}. We will come back to this point later.
It is unfortunately not possible to make a significant comparison with the
experimental data for these deformed nuclei because the experimental error
bars are much too large.

\section{Transition to zero pairing}
\label{transition}

Two types of calculations have been performed to study the OES in the zero pairing limit.
We have first used a schematic non self-consistent BCS model in order to get 
qualitative informations. Then, fully self-consistent HFB calculations 
have been performed in order to take into account rearrangement effects 
and obtain quantitative results.

\subsection{Schematic model presentation}
\label{subsecpresschem}

The schematic model consists in a non self-consistent BCS scheme on top of a fixed single-particle spectrum $\left\{ e_{k} \right\}$ for one kind of particles only. No two-body force is included: the orbital dependent gap is given as an initial parameter and is parametrized as:

\begin{eqnarray}
\Delta_k &=&  \Delta \, \, \, {\rm Exp} \left[ -\left(\frac{e_k-\lambda}{3\Delta}\right)^2\right]   \, \, \, \, \, \, \, \, \, \, \, \, \, \, \, \, ,
\end{eqnarray}
$\Delta$ being the input. Let us mention that $\Delta$ is the gap at the fermi energy.

The fixed spectrum can be either a priori constructed or taken from the self-consistent calculation of an even nucleus as a typical spectrum in a narrow region around that nucleus. In what follows, the subscripts HFB(E) are changed into BCS(E).

The fully paired part of the energy as defined by the first term of Eq.~\ref{defener} is that of a fully paired BCS vacuum:

\be
E^{BCSE}(N) = \sum_{k} \, v^{2}_{k} \, e_{k}  - \frac{1}{2} \, \sum_{k>0} \frac{\Delta^{2}_{k}}{\sqrt{\left(e_{k}-\lambda^{N}\right)^{2} + \Delta^{2}_{k}}}  \,  \, , \vspace{0.3cm}
\label{enerbcse}
\ee
$\lambda^{N}$ being fixed by the additional condition $< \hat{N} >$  = $N$ (odd or even). For $N$ even, it corresponds to the exact BCS energy. For odd $N$ the BCS energy is:

\be
E^{BCS}(N) = E^{BCSE}(N) + Min \, \left\{ E^{qp}_{k} \right\} \,  \, , \vspace{0.3cm}
\label{enerbcse}
\ee
where the $E^{qp}_{k}(N) = \sqrt{\left(e_{k}-\lambda^{N}\right)^{2} + \Delta^{2}_{k}}$ are the quasi-particle energies evaluated in the odd vacuum~\cite{dug3}. Once these energies
are given, the energy differences $\Delta^{(3)}_{BCS}$, $\Delta^{(3)}_{BCSE}$, $\Delta^{(5)}_{BCS}$ and $\Delta^{(5)}_{BCSE}$ can be computed.

\begin{figure}
\begin{center}
\leavevmode
\centerline{ \psfig{figure=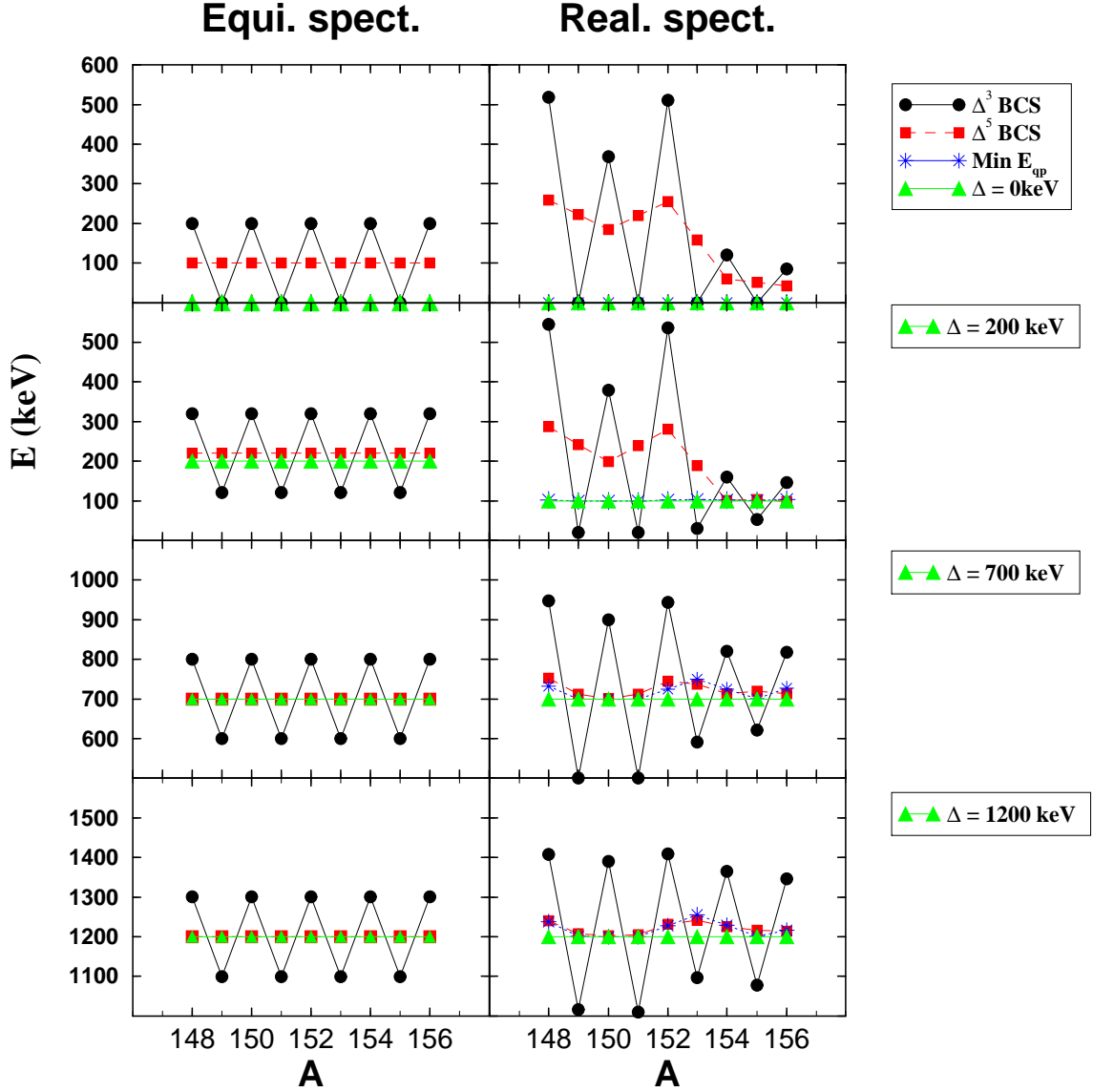,height=14cm}}
\end{center}
\caption{$\Delta^{(3)}_{BCS}$, $\Delta^{(5)}_{BCS}$ and $\Delta$ as a function of A. From top to bottom, the pairing gap increases from 0 to 1200 keV. Left column: calculation with an equidistant doubly degenerate spectrum. Right column: calculation using the self-consistent neutron HF spectrum of $^{152}$Ce. In addition to $\Delta$, the lowest qp in odd nuclei is shown. Results are displayed between $^{148}$Ce and $^{156}$Ce.}
\label{dtot_p_Ce_equi}
\end{figure}

\subsection{Results on cerium isotopes}
\label{subsecrescer}

We have performed two different applications of the schematic model.
First, we have used an equidistant two-fold degenerate spectrum simulating a deformed nucleus with a single-particle level spacing $\delta\epsilon$ = 400 keV. The calculation has been performed for six different values of $\Delta$, from zero to a typical value of 1200 keV. In the second case, we have used a realistic spectrum of $^{152}$Ce (self-consistent HF spectrum obtained with the SLy4 interaction). Fig.~\ref{dtot_p_Ce_equi} displays the evolution of $\Delta^{(3)}_{BCS}$, $\Delta^{(5)}_{BCS}$ with increasing gap $\Delta$ from $^{148}$Ce to $^{156}$Ce for both spectra. The left column is for the equidistant spectrum while the right one is for the realistic spectrum.

\begin{figure}
\begin{center}
\leavevmode
\centerline{ \psfig{figure=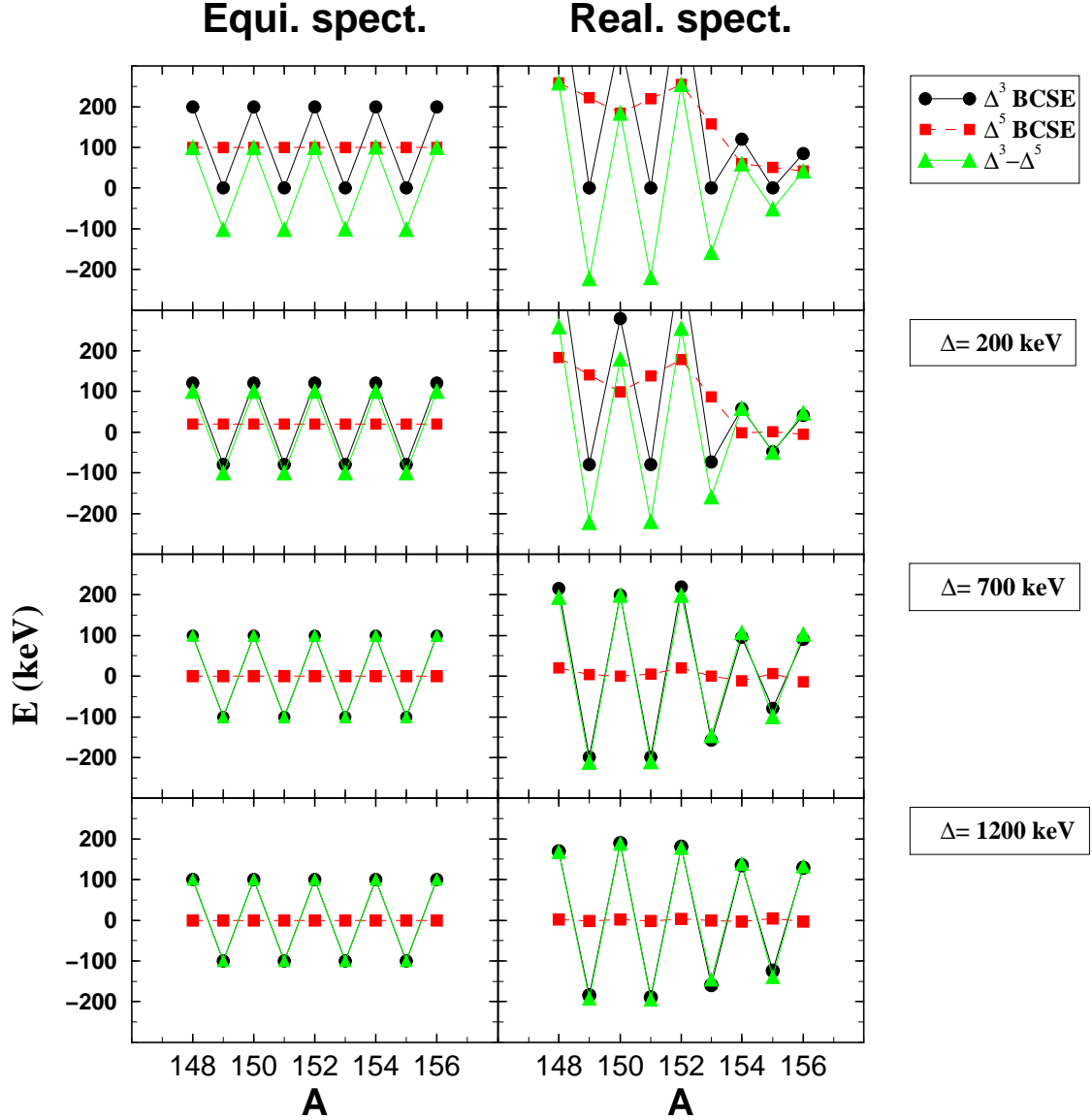,height=14cm} }
\end{center}
\caption{Energy differences $\Delta^{(3)}_{BCSE}$, $\Delta^{(5)}_{BCSE}$ and $\Delta^{(3)}_{BCS}$ - $\Delta^{(5)}_{BCS}$. Left and right panels: same as for Fig.~\ref{dtot_p_Ce_equi}. The results are displayed between $^{148}$Ce and $^{156}$Ce.}
\label{dp_ce_equ_sel}
\end{figure}

For $\Delta$ = 0, we recover qualitatively the results obtained in HF calculations without time-reversal symmetry breaking (cf. Eq.~\ref{delta3hf}). Namely, $\Delta^{(3)}_{BCS}(N)$ oscillates between 0 for odd $N$ and a non zero value for even $N$; $\Delta^{(5)}_{BCS}$ is different from zero for all values of $N$. It follows that in this case, $\Delta^{(3)}_{BCS}(odd)$ extracts the gap. 

When $\Delta$ is increased, this is no longer true. Odd-even mass differences are shifted to higher values in such a way that $\Delta^{(5)}_{BCS}$ tends to extract the pairing gap while $\Delta^{(3)}_{BCS}$ oscillates around this value. From the equidistant spectrum calculation, one can see that this asymptotic situation is achieved for a ratio $\Delta$/$\delta\epsilon$~$\approx$~0.5; i.e as soon as $\Delta$ reaches 200 keV. In real nuclei, apart for near closed-shell nuclei, typical $\Delta$/$\delta\epsilon$ values are greater than 0.5. 

The right hand side of Fig.~\ref{dtot_p_Ce_equi} shows that similar qualitative results are obtained using a realistic spectrum for deformed cerium isotopes, even though the structure of the realistic spectrum modifies the artificial regularities of the former case. The discrepancy between $\Delta$ and $Min \, \left\{ E^{qp}_{k} \right\}$ observed for some nuclei is due to the fact that the qp in the odd fully paired vacuum is not always such that $u^{2} - v^{2}$ is exactly 0~\cite{dug3}. From a quantitative point of view, $\Delta^{(5)}_{BCS}$ extracts precisely the quasi-particle energy as soon as $\Delta$ at the Fermi energy reaches 60 $\%$ of the realistic value obtained in an HFB calculation of these isotopes.

This can be understood from Fig.~\ref{dp_ce_equ_sel} which gives the E$^{BCSE}$ contribution to the different odd-even mass formulas for identical values of 
the gap, using both spectra.
 The oscillation of $\Delta^{(3)}_{BCS}(N)$ around 
$\Delta^{(5)}_{BCS}(N)$ is also plotted. For $\Delta$ = 0,  
$\Delta^{(3,5)}_{BCSE}$ is equal to $\Delta^{(3,5)}_{BCS}$ as in this case E$^{BCS}$ and E$^{BCSE}$ are the same for odd nuclei as well ($Min \, \left\{ E^{qp}_{k} \right\} = 0$ in the odd vacuum for a vanishing pairing). Through the even part of the energy E$^{BCSE}$, we have isolated the quantity responsible for the Jahn Teller OES of Eq. \ref{delta3hf} in absence of pairing. However, as the pairing increases this part of the energy becomes smoother with $N$, in such a way that $\Delta^{(5)}_{BCSE}$ goes to zero while $\Delta^{(3)}_{BCSE}$ oscillates regularly around it. When pairing increases, the deformation effect on the OES is equally redistributed to even and odd $\Delta^{(3)}(N)$ in such a way that it produces the staggering of $\Delta^{(3)}_{BCS}$ around $\Delta^{(5)}_{BCS}$ .

These schematic calculations help understanding the link between the apparently contradictory results obtained in HF calculations where $\Delta^{(5)}_{HF}$ is not zero and the BCS ones with the pairing turned on where $\Delta^{(5)}_{BCS}$ equals $\Delta_{pairing(+pol)}$. The contributions to $\Delta^{(5)}$ coming from the two parts of the energy evolve in opposite ways. The even contribution decreases from a non zero value because of the deformation effect described in Ref.~\cite{SDN98}, to zero value with increasing pairing. At the same time, the blocking contribution $\Delta_{pairing(+pol)}$ increases with pairing as expected. This illustrates why and how $\Delta^{(5)}$ extracts $\Delta_{pairing(+pol)}$ only, for a realistic pairing strength. 

\begin{figure}
\begin{center}
\leavevmode
\centerline{\psfig{figure=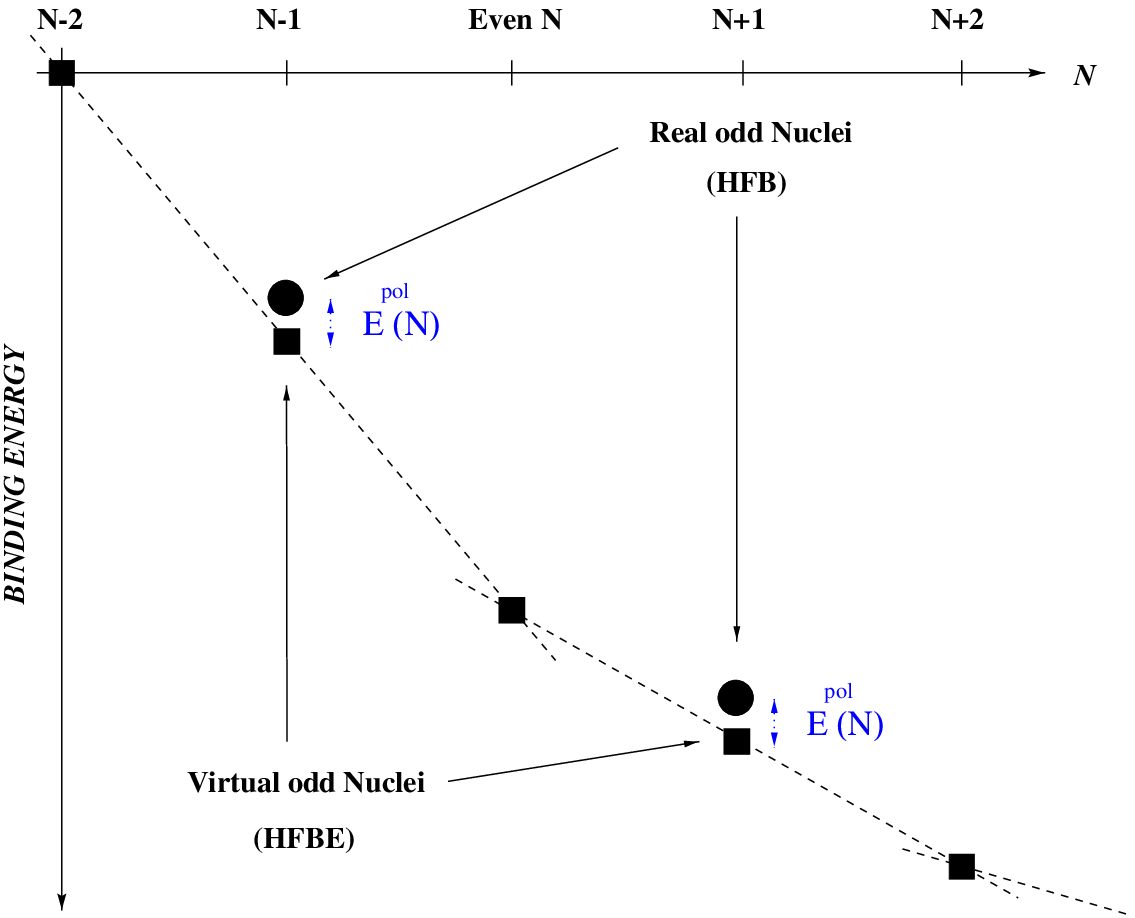,height=6cm} \psfig{figure=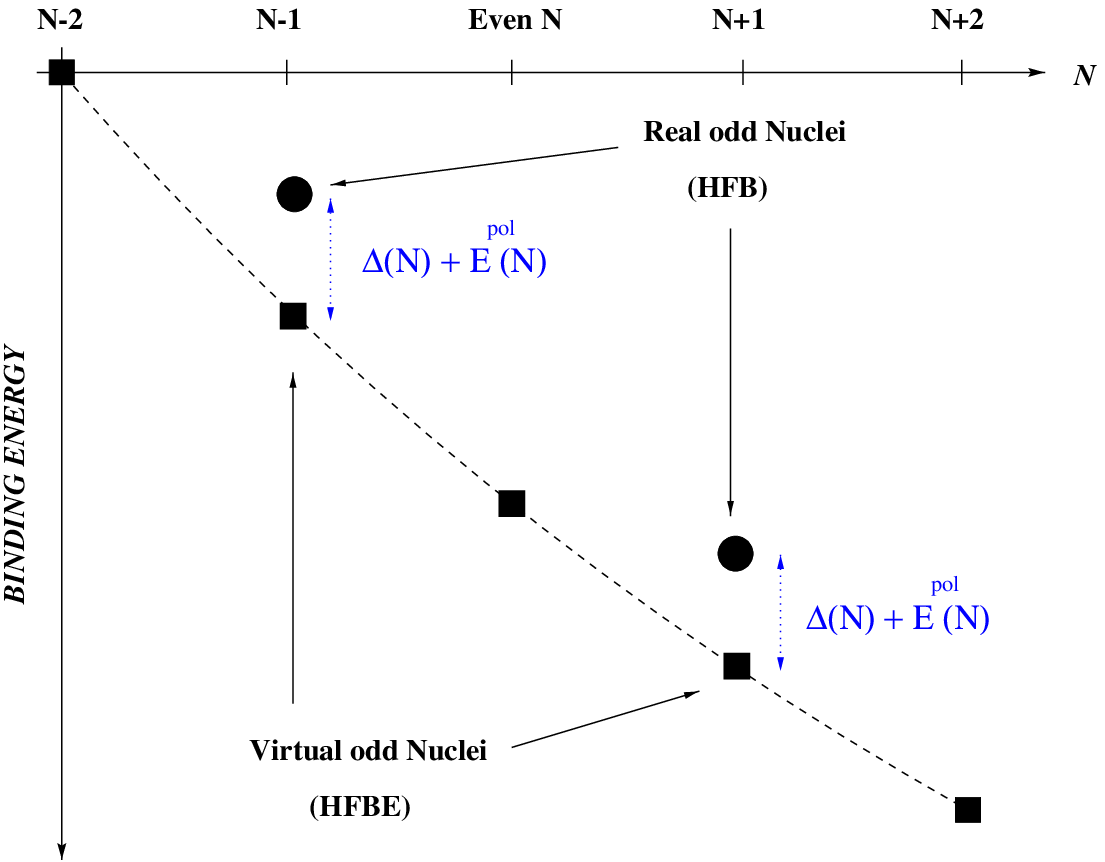,height=6cm}}
\end{center}
\caption{Binding energy as a function of N for the even part (squares joined by dashed line) and for the full odd states (circle). Left panel: no pairing. Right panel: realistic pairing case.}
\label{stagg10}
\end{figure}

Fig.~\ref{stagg10} gives a graphic representation of E$^{HFB}$ and E$^{HFBE}$ as a function of $N$ for both the zero pairing and the realistic pairing cases. It is drawn for an underlying doubly degenerate single-particle spectrum typical of a deformed nucleus. The left panel gives the zero pairing case where one can see that the even part of the energy is not a smooth function of $N$ and that the exact odd energy differs from E$^{HFBE}$ only if polarisation is included. The Kramers degeneracy is responsible for the linear behavior of E$^{HFBE}$ between two successive even nuclei while the single-particle level spacing is responsible for the change of the corresponding slope and induces the asymmetric Jahn Teller OES of Eq.~\ref{delta3hf}. In the realistic pairing case displayed on the right panel, the behavior of E$^{HFBE}$ becomes smooth with the nucleon number and no asymmetry between odd and even $N$ remains. As a consequence, one can graphically see that the Jahn  Teller OES cannot be transposed to the realistic pairing case as it is in the absence of pairing. 

\subsection{Results on tin isotopes}
\label{subsecrestin}

\begin{figure}
\begin{center}
\leavevmode
\centerline{ \psfig{figure=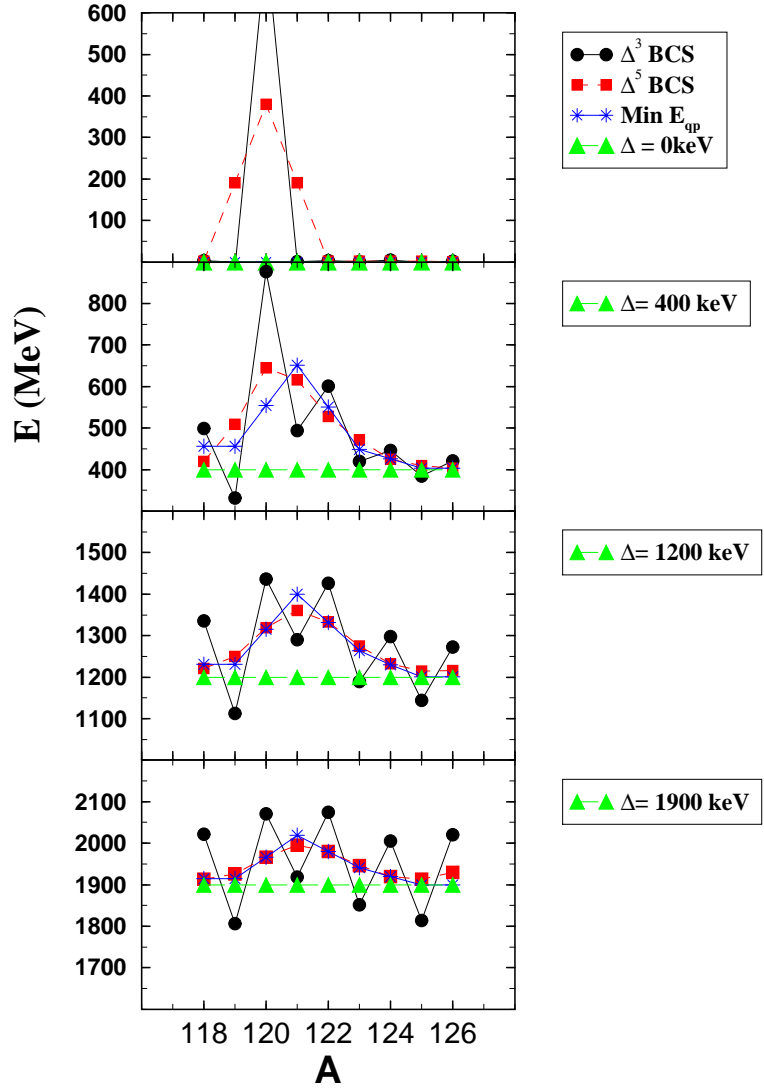,height=14cm}}
\end{center}
\caption{Same as Fig.\ref{dtot_p_Ce_equi} for tin isotopes. The calculation is done using $^{122}$Sn HF spectrum. Results are displayed between $^{118}$Sn and $^{126}$Sn.}
\label{dtot_Sn_self}
\end{figure}

For spherical nuclei,
the same kind of $\Delta^{(3)}$ staggering as for deformed nuclei is observed 
experimentally and found theoretically in HFB calculations, 
while  such staggering does not occur at the HF level because of the strong degeneracy of the spherical shells. 

We now apply our model with the HF spectrum of $^{122}$Sn. 
Fig.~\ref{dtot_Sn_self} displays the same quantities as 
Fig.~\ref{dtot_p_Ce_equi} between $^{118}$Sn and $^{126}$Sn, 
with a gap varying between 0 and 1900 keV. 
This latter value corresponds to the theoretical gap at the Fermi energy in 
the HFB calculations. 
As expected, no $\Delta^{(3)}_{BCS}$ staggering is seen for $\Delta$ equals 0 
apart for the transition between 2d3/2 and 1h11/2 spherical shells
which occurs at $N$ equals 120. 

As  $\Delta$ increases (from top to bottom on Fig.~\ref{dtot_Sn_self}), 
two modifications on odd-even differences occur simultaneously, namely the 
appearance of an odd-even staggering for $\Delta^{(3)}_{BCS}$ and the 
extraction of $Min \, \left\{ E^{qp}_{k} \right\}$ 
through $\Delta^{(5)}_{BCS}$. As for deformed nuclei, the  oscillating behavior of $\Delta^{(3)}_{BCS}$ is directly related to the contribution from the even part of the energy.
This transition from a situation where no $\Delta^{(3)}_{BCS}$ staggering 
exists for zero pairing to a situation where a clear staggering develops, 
shows that the calculations with and without pairing are not
related in a simple way.

\subsection{Single-particle level spacing and odd-even mass formulas}
\label{subsecsingpart}

\begin{figure}
\begin{center}
\leavevmode
\centerline{ \psfig{figure=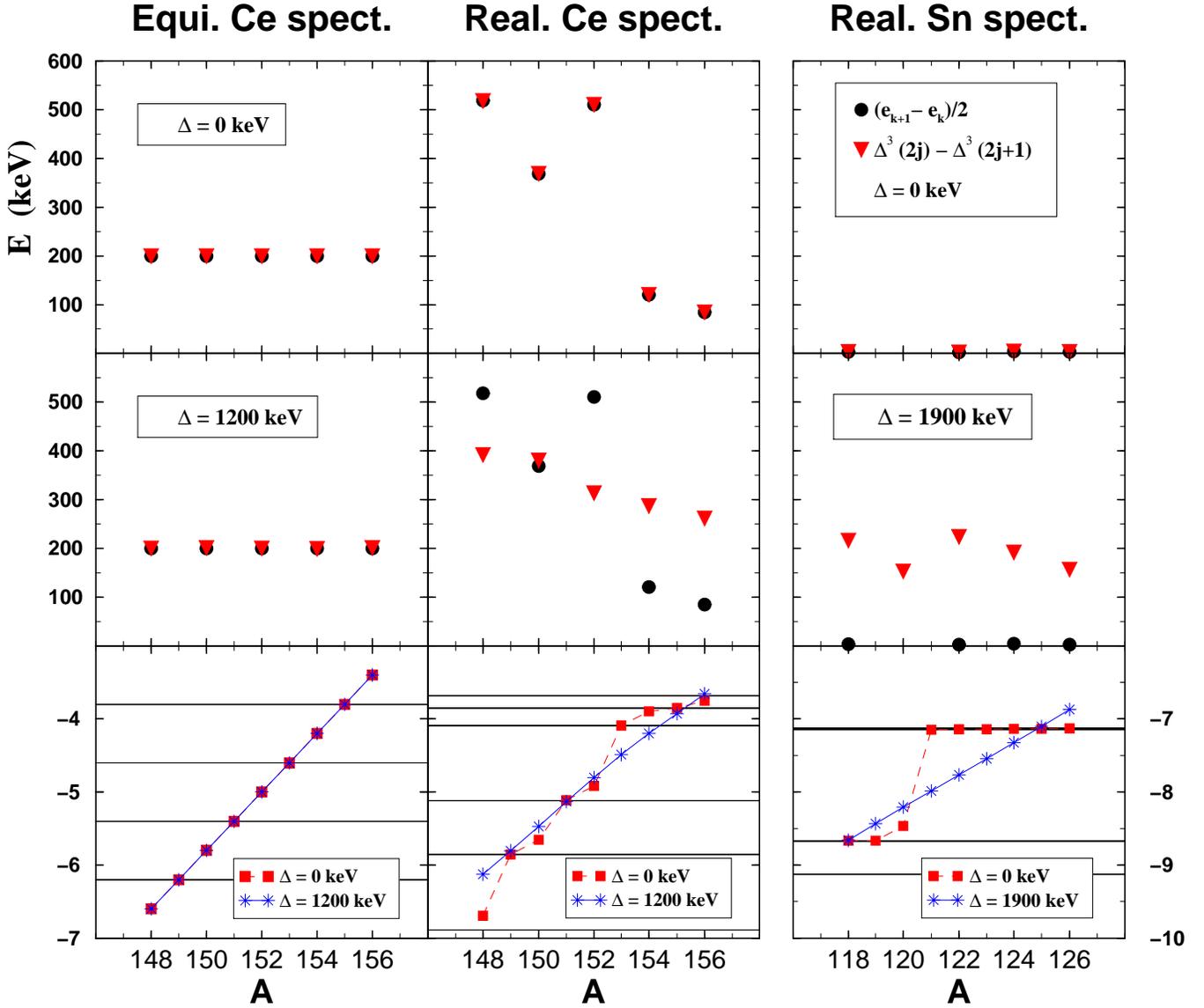,height=14cm}  }
\end{center}
\caption{Upper rows: $\Delta^{(3)}(2j)$ - $\Delta^{(3)}(2j+1)$ compared to the splitting around the Fermi level in the even nucleus ($e_{k+1}$ - $e_{k}$)/2 for two extreme values of the gap (top, $\Delta$ = 0; middle, $\Delta_{realistic}$). The 1.5 MeV splitting for $^{120}$Sn is out of scale. Lower row: chemical potential in MeV as a function of $A$ for the two values of $\Delta$.}
\label{diffsingpart_ce_sn}
\end{figure}

In Ref.~\cite{SDN98,doba}, it has been suggested that the $\Delta^{(3)}$ 
oscillation as a function of N 
could be used as a measure of the splitting of the 
single-particle spectrum around the Fermi level for even deformed 
nuclei (see Eq.~\ref{delta3hf}). 
To study the validity of this statement, we compare on
Fig.~\ref{diffsingpart_ce_sn} 
$\Delta^{(3)}_{BCS}(2j)$ - $\Delta^{(3)}_{BCS}(2j+1)$ and 
($e_{k+1}$ - $e_{k}$)/2 for zero and a realistic value of
the pairing strength. The variation of the chemical potential 
as a function of $A$ is also plotted. 
In the left column are presented the results of the calculations 
for the cerium isotopes with an equidistant spectrum 
and with a realistic spectrum in the middle. In the
right part of the figure are shown the results for tin isotopes using 
a realistic spectrum. 

For $\Delta$ = 0 and 1200 keV, the splitting of the single-particle energies
in the cerium spectrum 
is exactly reproduced by the staggering of $\Delta^{(3)}$ 
if the equidistant spectrum is used. 
This result suggests that one can extract informations 
about the neutron (proton) single-particle spectrum 
along an isotopic (isotonic) line through odd-even mass differences 
whether pairing is present or not. 
On the other hand, using a realistic spectrum this conclusion is valid 
only in the limit of vanishing pairing. 
For a realistic value of $\Delta$, the difference 
$\Delta^{(3)}_{BCS}(2j)$ - $\Delta^{(3)}_{BCS}(2j+1)$ 
is no longer a measure of ($e_{k+1}$ - $e_{k}$)/2. 
This is actually the case as soon as $\Delta$ reaches 60$\%$
of a realistic value. 
This is further confirmed by the calculation on tin isotopes where 
$\Delta^{(3)}_{BCS}(2j)$ - $\Delta^{(3)}_{BCS}(2j+1)$ is non zero for 
realistic pairing strengths whereas the corresponding 
spherical single-particle energies are highly degenerate (see the 
 right column).

This result can be understood as a consequence of the very different
way a nucleon is added whether pairing correlations are present or not.
When starting from an HF state, the pairing is increased, 
the amount of binding energy associated with 
the addition of a nucleon in the even structure is less and less 
related to a specific single-particle energy.
Rather, the nucleon is  spread out on the levels around the Fermi level 
because of 
pairs scattering~\cite{dug3}. Consequently, the memory of the underlying 
single-particle spectrum is washed out. Besides, this is the reason why the HFBE energy becomes smoother as a function of $A$ with increasing pairing (cf. Fig.~\ref{stagg10}).

The bottom row of Fig.~\ref{diffsingpart_ce_sn} illustrates 
the previous statement by showing the chemical potential 
in the even state as a function of the mass number. 
Let us consider the two cases which make use of 
a realistic spectrum (middle and right panels of the bottom row). 
For $\Delta$ = 0, $\lambda$ is sensitive to the orbits, 
while for $\Delta_{realistic}$ it behaves more smoothly as a 
function of $A$ and does not 
reflect the structure of the spectrum anymore. 
Thus, since one has

\begin{equation}
\Delta^{(3)}_{BCS}(2j)\!-\!\Delta^{(3)}_{BCS}(2j+1) \!
\approx \! \frac{\partial^{2} E^{BCSE}}{\partial N^{2}} \! \approx \!
 \frac{\partial \lambda}{\partial N}
\label{derivelambda}
\end{equation}
\vspace{0.1cm}

\noindent this observable is no longer 
directly related to the single-particle level spacing around the Fermi level 
for a realistic pairing strength. 
The strong influence of the single-particle levels structure is lost for a value of the gap smaller than typical splittings in the spectrum ($\Delta \epsilon$ goes from 200 keV to 1 MeV in the studied cerium region).

In the calculation based on an equidistant spectrum, 
the left panel of the bottom row illustrates  why in this case
one can still extract
informations about the single-particle level spacings through 
odd-even mass differences for a realistic pairing intensity. The evolution of $\lambda$ with A does not depend on
$\Delta$. Indeed, even if the nucleon is spread over the Fermi sea, 
the average effect of the pair scattering process cancels 
out because of the symmetry of this spectrum. 
As a result, the  energy added by the extra nucleons
remains equal to the single-particle 
energy of the orbit on which the nucleon is put in absence of pairing. 
This points out the inherent limits of schematic models used with 
very simplified single-particle spectra.

The above result could not have been worked out for spherical nuclei with models limited to a single $j$ shell. Indeed, from the bottom row of Fig.~\ref{diffsingpart_ce_sn}, one sees that the 
effect involves several spherical shells for the pair scattering. For instance, the pair scattering effect is efficient enough for $\Delta$ equal to $\Delta_{realistic}$ to loose the information about the 1.5 MeV splitting between the 2d3/2 and 1h11/2 spherical shells.

\subsection{Self-consistent calculations}
\label{subsecresself}

\begin{figure}
\begin{center}
\leavevmode
\centerline{ \psfig{figure=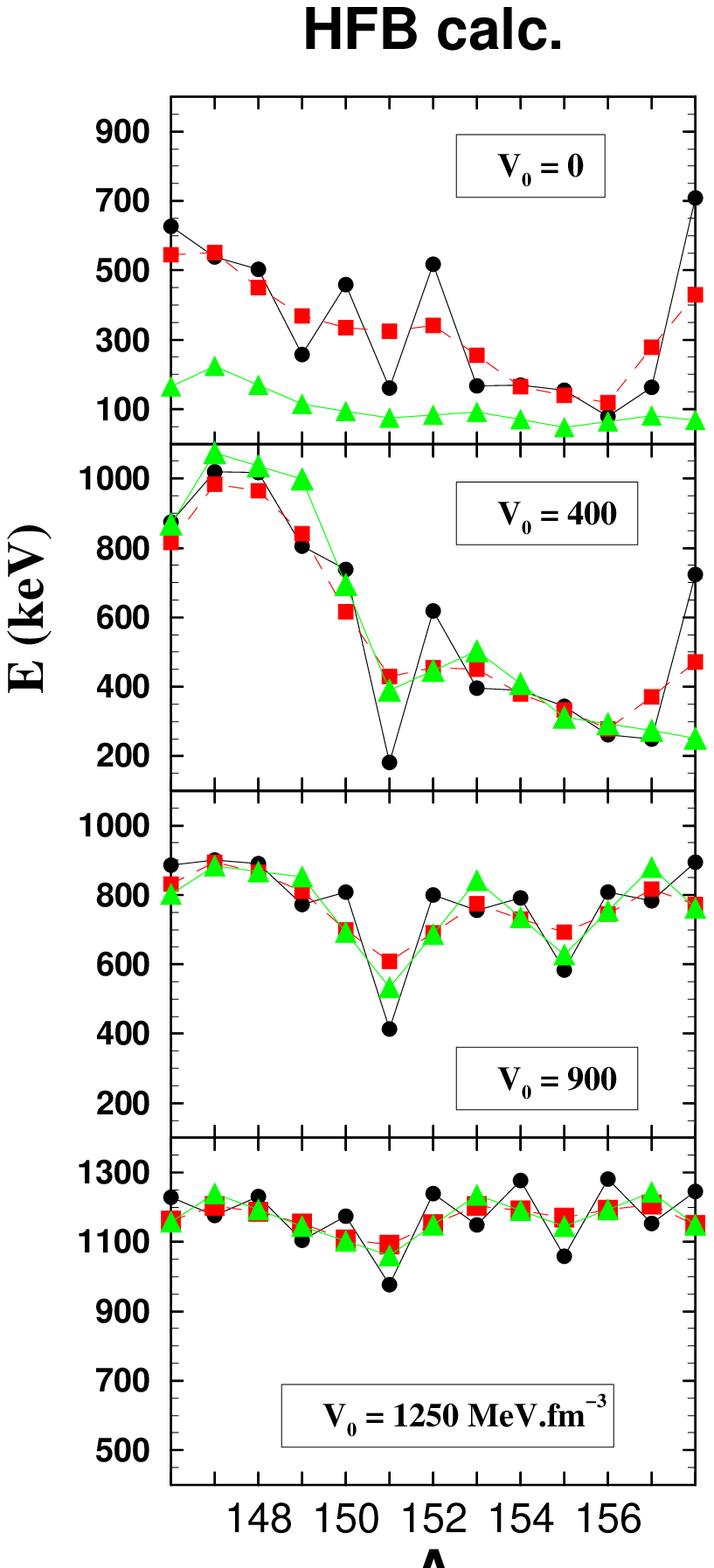,height=15cm} \psfig{figure=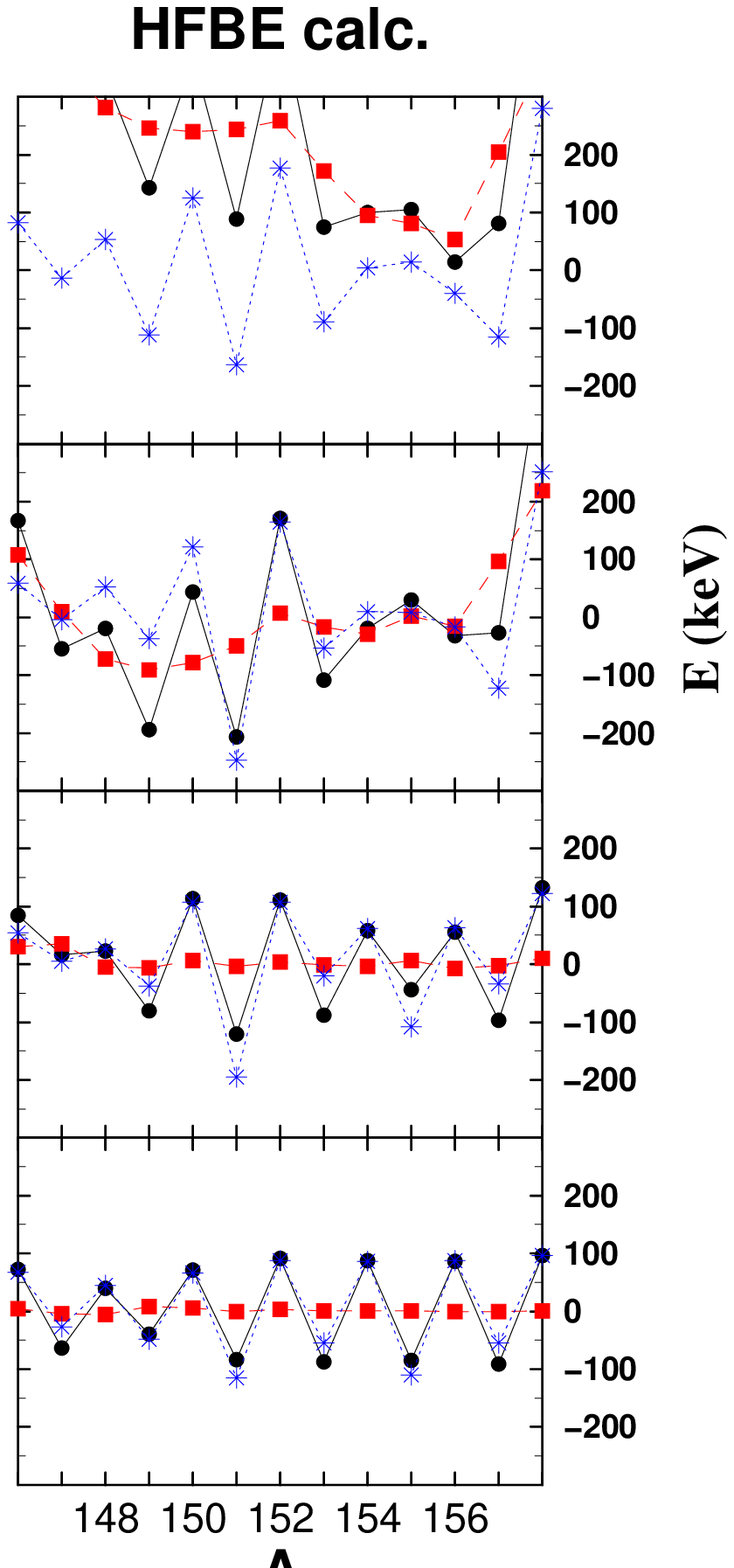,height=15cm} }
\end{center}
\caption{Left (right) column: $\Delta^{(3)}_{HFB(E)}$ (circles with solid line), $\Delta^{(5)}_{HFB(E)}$ (squares with dashed line), self-consistent qp energy E$^{HFB}$ - E$^{HFBE}$ (triangles with dotted line) and $\Delta^{(3)}_{HFB} - \Delta^{(5)}_{HFB}$ (stars with dotted line). From top to bottom the intensity of the neutron pairing force increases from 0 to 1250 MeV.fm$^{-3}$. Results are displayed between $^{148}$Ce and $^{156}$Ce.}
\label{dtot_p_Ce}
\end{figure}

Let us present the same analysis for fully self-consistent HFB calculations of cerium isotopes. Pairing correlations are gradually turned on through the increase of the neutron pairing force intensity $V_{n}$ up to the realistic case presented in section~\ref{subsecresdef}. Fig. \ref{dtot_p_Ce} displays the same quantities as Fig.~\ref{dtot_p_Ce_equi} and~\ref{dp_ce_equ_sel}. Instead of the perturbative BCS quasi-particle energy $Min \, \left\{ E^{qp}_{k} \right\}\,$, the energy difference E$^{HFB}$ - E$^{HFBE}$ is given. This quantity is the self-consistent version of the created quasi-particle energy in odd nuclei~\cite{dug3}. The upper left panel shows that it is non-zero in the zero pairing case since it already contains the time-reversal symmetry breaking effect which shifts up all the $\Delta^{(n)}_{HFB}$. 

First, let us concentrate on the odd-even differences of E$^{HFBE}$ for $V_{n} = 0$ (right upper panel). For this particular case, we will use HF and HFE subscripts instead of HFB and HFBE. There are neither polarisation nor pairing effects and one can focus on the strong influence on the results of the self-consistency in the mean-field treatment. Contrary to the schematic results, $\Delta^{(3)}_{HFE}(odd)$ can be significantly different from zero in this case (several hundreds keV for example between $^{145}$Ce and $^{149}$Ce). Self-consistency significantly modifies the picture as compared to the independent particle scheme, especially in regions of varying deformation~\cite{SDN98,flo}. 
Since $\Delta^{(3)}_{HFE}(odd)$ is not 0, it is thus difficult to argue that it extracts $\Delta_{pairing}$. The only possible statement is that in the zero pairing case an OES is seen with an oscillating $\Delta^{(3)}_{HF}$ together with a non-zero $\Delta^{(5)}_{HF}$. 

The energy difference $\Delta^{(3)}_{HF}(odd)$ is 
closer to E$^{HF}$ - E$^{HFE}$ than $\Delta^{(5)}_{HF}$. 
However, as the pairing intensity increases $\Delta^{(5)}_{HFB}$ 
extracts the energy difference E$^{HFB}$ - E$^{HFBE}$, which is nothing 
but the staggering of the energy associated with blocking, while $\Delta^{(3)}_{HFB}$ 
oscillates around this value. 
This is the case as soon as the pairing intensity 
reaches about 72 $\%$ of the realistic value 
($V_{n}$ = $-$ 900 MeV.fm$^{-3}$). This statement is valid 
even when self-consistency effects are large 
in this region of varying deformation. The results are presented 
only for a small part of the cerium isotopic line, the same conclusions 
holds for the whole line. 

From the right column giving the even contributions for increasing strength $V_{n}$, one can see that the analysis of the schematic calculations remain true. Namely, as also discussed in section~\ref{subsecresdef}, the oscillations of $\Delta^{(3)}_{HFB}$ around $\Delta^{(5)}_{HFB}$ are reproduced 
quantitatively by the contribution from the even, smooth, part of 
the energy for realistic $V_{n}$.

\begin{figure}
\begin{center}
\leavevmode
\centerline{ \psfig{figure=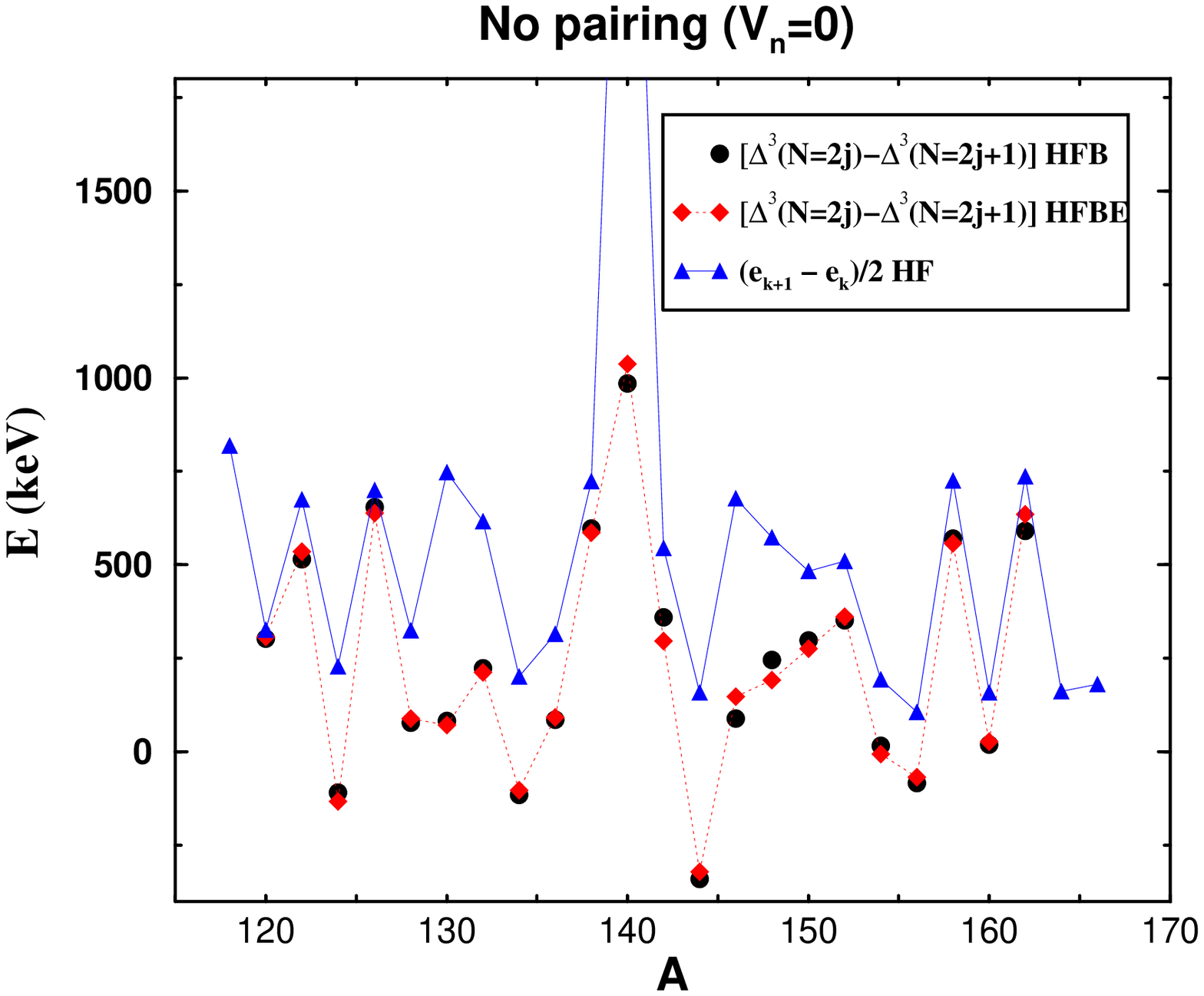,height=6cm} \psfig{figure=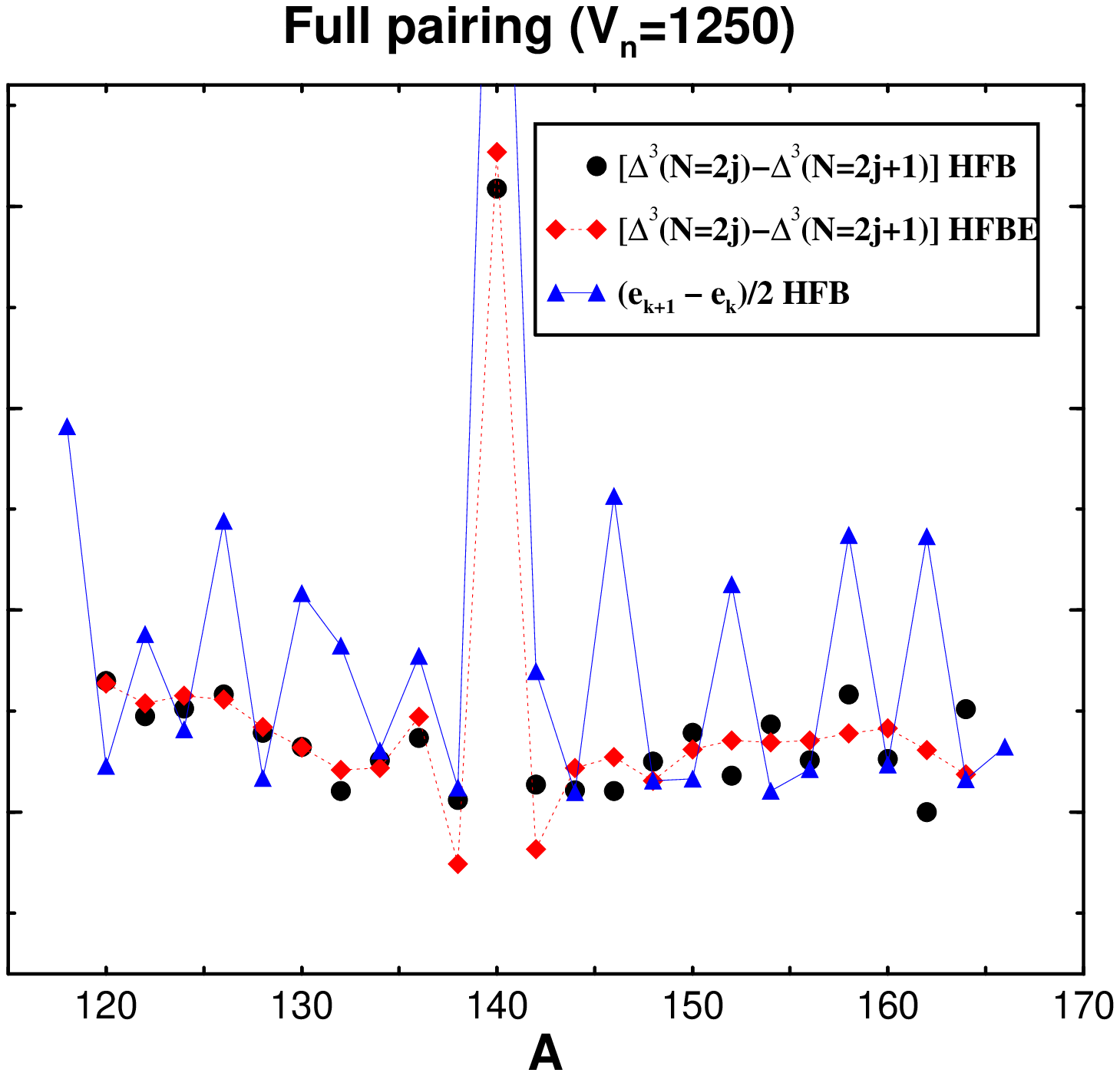,height=6cm} }
\end{center}
\caption{Energy differences $\Delta^{(3)}_{HFB}(2j) - \Delta^{(3)}_{HFB}(2j+1)$, $\Delta^{(3)}_{HFBE}(2j) - \Delta^{(3)}_{HFBE}(2j+1)$ and single-particle splitting $(e_{k+1} - e_{k})/2$ around the Fermi energy for self-consistent calculations of cerium isotopes. }
\label{derniere}
\end{figure}

On Fig.~\ref{derniere}, the  $\Delta^{(3)}_{HFB}$ and 
$\Delta^{(3)}_{HFBE}$ staggerings are compared
with the single-particle level spacing 
around the Fermi energy in even nuclei. 
The left panel displays $\Delta^{(3)}_{HFB}(2j) - \Delta^{(3)}_{HFB}(2j+1)$,
 $\Delta^{(3)}_{HFBE}(2j) - \Delta^{(3)}_{HFBE}(2j+1)$ and 
$(e_{k+1} - e_{k})/2$ for $V_{n} = 0$. 
The staggerings of $\Delta^{(3)}_{HFB}$ and
$\Delta^{(3)}_{HFBE}$ coincides. 
It means that this staggering is entirely due to the even contribution
to the EOS. 
Moreover, the $\Delta^{(3)}$ staggering roughly
extracts the splitting in the HF spectrum except in the region of 
varying deformation ($^{124}$Ce to $^{152}$Ce) 
where the  rearrangement due to self-consistency from one nucleus to the next
is large.

The right panel displays the same quantities in the case of realistic pairing intensity. With pairing included, $e^{V_{n} \neq 0}_{k}$ is the eigenenergy of the HF field deduced from $e^{V_{n}=0}_{k}$ by continuity. Again, the $\Delta^{(3)}_{HFB}$ staggering is well reproduced by 
$\Delta^{(3)}_{HFBE}(2j) - \Delta^{(3)}_{HFBE}(2j+1)$ along the whole line, 
whatever the magnitude of the self-consistency effects is. 
However, the information about the HF eigen-energies is lost in this case. Indeed, the addition of a nucleon is no longer related 
to a single orbit when pairing is included~\cite{dug3}.  
These conclusions are the same as in the case of the schematic model.

\section{Analysis and conclusions}
\label{secconclu}

We have proposed an analysis for the odd-even mass staggering based on the definition of a ``virtual'' odd nucleus (HFBE state) having the structure of an even one as the underlying structure of the ``real'' odd nucleus~\cite{bend,dug3}.

For realistic pairing intensities, it has been shown that the $\Delta^{(5)}(N)$ mass formula extracts precisely the self-consistent HFB quasi-particle energy for spherical as well as for deformed nuclei. The self-consistent HFB quasi-particle energy corresponds to the blocking of the odd nucleon on top of the fully paired odd reference vacuum and contains both the pairing gap, $\Delta(N)$, and the time-reversal symmetry breaking effect, E$^{pol}$.

Similar results have already been reported for spherical nuclei in Ref.~\cite{bend} where the extraction through $\Delta^{(5)}$ of the pure blocking contribution to the ground-state energy of odd nuclei was  pointed out. However, this work was done in the HF + BCS framework and without breaking time-reversal symmetry and $\Delta^{(5)}$  extracts then only the self-consistent pairing gap (E$^{pol}$ = 0). Such an approximation limits the pertinence of the comparison with experimental data. Our present study does incorporate this physical effect and extends that earlier work to realistic cases and to deformed nuclei. 

Satula {\it et al.}~\cite{SDN98} made a similar study for deformed nuclei 
in the HF approximation, as a reference to identify pairing contributions to the OES. They proposed to use $\Delta^{(3)}(odd)$ as a measure of the pairing gap and the difference $\Delta^{(3)}(even)$ - $\Delta^{(3)}(odd)$ as the Jahn Teller contribution to the OES (contribution from deformation). They have also neglected time-reversal symmetry breaking effects. 

In this context of time-reversal invariance, our extended analysis of the OES as a function of pairing correlations intensity within the frame of a schematic BCS model has allowed to sort out the contradictory former propositions. The conclusions of Satula {\it et al.}, based on Eq.~\ref{delta3hf}, have been shown to hold only for very weak pairing. For a realistic pairing intensity, $\Delta^{(3)}(odd)$ is no longer a measure of the gap alone since it contains an additional contribution coming from the even part of the energy E$^{HFBE}$ as defined in Eq.~\ref{defener} (see section~\ref{transition}). On the other hand, $\Delta^{(5)}$  extracts the pairing gap in this case. Self-consistent HFB calculations have confirmed these conclusions from a quantitative point of view.

Let us go one step further by introducing, in a self-consistent mean-field picture, the time-reversal symmetry breaking effect on binding energy. This effect is formally related to the physical blocking process in odd nuclei as extensively discussed in Ref.~\cite{dug3}. It follows that it is deeply associated with the self-consistent pairing gap in such a way that these two energetic quantities cannot be separated through odd-even mass differences. They are both contained into $\Delta^{(n)}(N)$ at all orders in $n$ when using experimental data. Consequently, one has to include this effect in realistic calculations in order to compare directly theoretical and experimental odd-even mass differences.

Finally, we have identified in the present paper the physical content of $\Delta^{(3)}$ and $\Delta^{(5)}$ in fully self-consistent mean-field calculations including realistic pairing:

\begin{eqnarray}
\Delta^{(3)}_{HFB}(N) \, &\approx& \, \Delta(N) \, + \,  E^{pol}(N) + \frac{(-1)^{N}}{2} \, \left. \frac{\partial^{2} E^{HFBE}}{\partial N^{2}} \, \right|_{N}                 \label{conclu1}   \\
\Delta^{(5)}_{HFB}(N) \, &\approx& \, \Delta(N) \, + \,  E^{pol}(N)  \label{conclu2}  \, .
\end{eqnarray}
\vspace{0.4cm}

\noindent where in the picture of Ref.~\cite{dug3}, $1/2 \,\partial^{2} E^{HFBE}/\partial N^{2}$ is related to the nucleon addition process and contains the full assymetry energy contribution to the OES whereas $\Delta(N) +  E^{pol}(N)$ is related to the blocking of this nucleon.

Comparing their results with those obtained by 
Satula {\it et al.}~\cite{SDN98}, Bender {\it et al.}~\cite{bend} 
argue that the Jahn Teller effect  (called ``mean-field effect'' since 
it is related to the structure of the single-particle spectrum) 
is not connected to the oscillation of $\Delta^{(3)}$ found 
for spherical nuclei (E$^{HFBE}$ contribution). 
We have demonstrated that the $\Delta^{(3)}$ staggering was always 
related to the E$^{HFBE}$ contribution (typically $\pm$50/150 keV). 
This energy is related in some way to the single-particle structure of 
a given nucleus, but our  extended schematic and fully self-consistent 
calculations have shown that 
the experimental $\Delta^{(3)}$ staggering cannot be identified 
with single-particle level spacing at the Fermi surface 
as suggested in Ref.~\cite{SDN98,doba}, apart for nuclei immediately near 
magic ones.

Our results are based on calculations done in the A = 100-170 mass region. 
They should also be valid for lighter nuclei. 
Indeed, the regime (independent particle or correlated system) 
in which the system stands depends on a typical ratio 
$\Delta$/$\delta\epsilon$. It has been shown in a schematic BCS model 
that the correlated regime is achieved for a 
ratio $\Delta$/$\delta\epsilon$~$\approx$~0.5
whereas for realistic calculations 
in the $A$ = 100-170 mass region, it is achieved for a value of the pairing gap a few
times smaller than the level spacings near the Fermi energy. 
These two arguments are in favor of the correlated limit for nuclei in the mass region A = 30-100 where $\Delta_{F}$/$\delta\epsilon_F$ is typically between 0.5 and 1 for mid-shell nuclei. In order to check this statement, we have performed an exploratory calculation for Mg isotopes between $^{24}$Mg and $^{28}$Mg using our schematic BCS model. 
 The single-particle spectrum and gap value at the Fermi energy taken from a self-consistent HFB calculation of $^{24}$Mg were used. 
The results support the extrapolation of our 
results to lighter masses. It also shows that the E$^{HFBE}$ contribution 
to $\Delta^{(3)}$ increases in average with decreasing mass number as the mean single-particle level spacing increases at the same time. It qualitatively explains the well known increase of the $\Delta^{(3)}$ staggering around $\Delta^{(5)}$ in light nuclei (see Fig. 3 of Ref.~\cite{SDN98} for instance).

The only limitation of the above conclusions concerns nuclei with neutron 
(proton) number one or two units away from magic numbers. 
These nuclei belong to the intermediate regime 
where $\Delta^{(3)}(odd)$ is of the same quality or better than $\Delta^{(5)}$
to extract informations about the blocking effect 
(see upper panels of Fig.~\ref{dtot_p_Ce_equi}). However, the limitation concerns a very limited number of nuclei which, in any case, should not be used for a study intended to adjust the pairing force. It has also been shown that the pairing force should be fitted on global observables such as rotation, and orbit-related observables as the OES in order to adjust at the same time the parts of the pairing energy contained in E$^{HFBE}$ and in $\Delta(N)$.

The above analysis is directly related to the nucleon addition process which is significantly modified by the inclusion of pairing correlations in the nuclear wave-function~\cite{dug3}. Besides, in a recent lecture~\cite{flo} where Satula and coworkers results on the OES were reported, Flocard suggested that it is somewhat surprising that the prescription of Eq.~\ref{delta3hf} derived using an independent particle picture remains correct for strongly correlated systems as nuclei. The present work has shown that this doubt was justified since pairing in such systems is strong enough in general to modify the picture by washing out the decisive influence of single-particle energies on odd-even effects. 

\begin{figure}
\begin{center}
\leavevmode
\centerline{ \psfig{figure=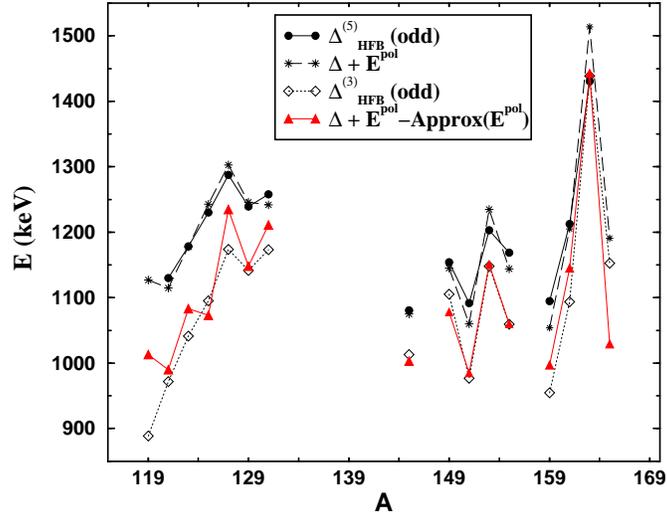,height=7cm} }
\end{center}
\caption{Comparison between $\Delta^{(5)}_{HFB}(odd)$ and the self-consistent qp energy $\Delta + E^{pol}$ along the cerium isotopic chain. The comparison between $\Delta^{(3)}_{HFB}(odd)$ and an approximation of the self-consistent pairing gap $\Delta$ is also given.}
\label{choix}
\end{figure}

Once we have identified the physical content of the $\Delta^{(3)}$ and $\Delta^{(5)}$ odd-even mass formulas, their respective advantages and drawbacks as suited quantities for a detailed study or fit of a pairing force remain to be analyzed. Eq.~\ref{conclu2}, shows that $\Delta^{(5)}$ contains one quantity in addition to the pairing gap. Actually, $\Delta^{(5)}(odd)$ for example contains a weighted average of $\Delta + E^{pol}$ over three odd nuclei. As shown on Fig.~\ref{choix}, this is responsible for a slight deterioration of the validity of identity \ref{conclu2} when $\Delta + E^{pol}$ changes suddenly around one nucleus (see $^{163}$Ce for instance). On the other hand, $\Delta^{(3)}(odd)$ contains $\Delta + E^{pol}$ from the studied nucleus only which is an advantage over $\Delta^{(5)}$. 

Eq.~\ref{conclu1} shows that $\Delta^{(3)}$ contains an extra contribution coming from the smooth part of the energy E$^{HFBE}$. In section~\ref{secres}, this extra contribution has been shown to be of the order of $\pm$50 to $\pm$100 keV in spherical tin isotopes and of the order of $\pm$100 to $\pm$150 keV in cerium deformed nuclei, namely it contributes for about 8 to 12 $\%$. 
Then, the time-reversal symmetry effect has been theoretically extracted in Ref.~\cite{dug3} through a perturbative calculation (labeled Approx(E$^{pol}$) in the present work) 
for the cerium isotopes and appeared to be of the order of +100 to +150 keV. 

It follows that the two last previous contributions roughly cancel out in $\Delta^{(3)}(odd)$ and that the relative weight of $\Delta(N)$ is larger in $\Delta^{(3)}(odd)$ than in $\Delta^{(5)}(odd)$. However, the details of this cancelation is not under control since
 $\Delta^{(3)}_{HFBE}$ and above all $E^{pol}(N)$, are not well known. In particular, the 
time-reversal symmetry breaking process deserves more studies since the results are force and model dependent~\cite{,dug3,satupol,rutz2}. In order to exemplify the situation, Fig.~\ref{choix} also gives $\Delta^{(3)}(odd)$ and an approximation of the self-consistent pairing gap, $[\Delta + E^{pol}$ - Approx$(E^{pol})]$. Results are not shown when the hypothesis of the perturbative calculation are not fulfilled~\cite{dug3}. One can see that $\Delta^{(3)}_{HFB}(odd)$ is often closer to the self-consistent pairing gap than $\Delta^{(5)}_{HFB}(odd)$ which means that the cancelation between the two different effects $E^{pol}$ and $\Delta^{(3)}_{HFBE}$ is quite effective in the present case. 

Finally, one should propose $\Delta^{(3)}(odd)$ as the better suited quantity for a detailed study of the pairing gap or for the fit of a pairing force through the adjustment of theoretical and experimental odd-even mass differences. We would like to stress the fact that this conclusion is not a validation of the analysis performed in~\cite{SDN98} as the way to reach it has been very different and needed the inclusion of time-reversal symmetry breaking in order to point out the a priori unexpected cancelation between $E^{pol}$ and $\Delta^{(3)}_{HFBE}(odd)$. Moreover, this conclusion still depends on more extensive analysis of the time-reversal symmetry breaking contribution in different mass regions to be done in order to study the presently found cancelation effect. As an example, we may interpret this effect to be responsible for the overall agreement found between  $\Delta^{(3)}_{exp}$ and several averaged\footnote{It consists of averaging over several isotones (isotopes)
 the neutron (proton) pairing gap in even nuclei, after a possible weighted average over the Fermi sea.} theoretical HFB pairing gaps in an extensive re-analysis of the commonly accepted formula $\Delta$ = 12 $A^{-1/2}$ MeV for the pairing gap as a function of the mass number~\cite{hilaire}.

It is important to stress that our purpose takes into account only one kind of pairing correlations, i.e. proton-proton and neutron-neutron pairing. The questions related to proton-neutron cooper pairs around $N$ = $Z$ nuclei need of course an extension of our approach. Satula and Wyss~\cite{wyss1}, Vogel~\cite{vogel} as well Terasaki {\it et al.}~\cite{tera2} have investigated these questions. Their conclusions correlated to an extension of our work could deliver a good indicator to fix the theoretical intensity of this neutron-proton pairing.

\section{Acknowledgment}
\label{secremer}

We would like to thank J. Dobaczewski for fruitful discussions.

%Bibliographie


\begin{thebibliography}{99}

\bibitem% [Boh8] 
          {Boh8}
A. Bohr, B.R. Mottelson and D. Pines, Phys. Rev. {\bf 110} (1958) 936

\bibitem% [BM75]
          {BM75}  
A. Bohr and B.R. Mottelson, Nuclear Structure (Benjamin, New York 1969) Vol. 1 

\bibitem% [jensen]
	  {jensen}
A.S. Jensen, P.G. Hansen and B. Jonson, Nucl. Phys. {\bf A431} (1984) 393

\bibitem% [mad]
          {mad}
D.G. Madland and J.R. Nix, Nucl. Phys. {\bf A476} (1988) 1

\bibitem% [mol]
          {mol}  
P. Moller and J.R. Nix, Nucl. Phys. {\bf A536} (1992) 20

\bibitem% [Haa98]
          {Haa98}  
H. H\"{a}kkinen, J. Kolehmainen, M. Koskinen, P.O. Lipas and M. Manninen,  Phys. Rev. Letters {\bf 78} (1997) 1034

\bibitem% [cle]
          {cle}  
K. Clemenger, Phys. Rev.  {\bf B32} (1985) 1359

\bibitem% [man]
          {man}  
M. Manninen, J. Mansikka-aho, H. Nishioka and Y. Takahashi, Z. Phys.  {\bf D31} (1994) 259; C. Yannouleas and U. Landman, Phys. Rev. {\bf B51} (1995) 190

\bibitem% [SDN98]
          {SDN98}
W. Satula, J. Dobaczewski and W. Nazarewicz, Phys. Rev. Letters {\bf 81} (1998) 3599

\bibitem% [bend]
	  {bend}
M. Bender, K. Rutz, P.-G. Reinhard and J.A. Maruhn, Eur. Phys. J. {\bf A8} (2000) 59

\bibitem% [dug3]
	  {dug3}
T. Duguet, P. Bonche, P.-H. Heenen and J. Meyer, article1   (2001) 

\bibitem% [doba2]
          {doba2}
J. Dobaczewski and J. Dudek, Phys. Rev. {\bf C52} (1995) 1827

\bibitem% [chab]
          {chab}
E. Chabanat, P. Bonche, P. Haensel, J. Meyer and R. Schaeffer,
Nucl. Phys. {\bf A627} (1997) 710; Nucl. Phys. {\bf A635} (1998) 231

\bibitem% [rutz]
	  {rutz}
K. Rutz, M. Bender, P.-G. Reinhard, J.A. Maruhn and W. Greiner, Nucl. Phys. {\bf A634} (1998) 67

\bibitem%  [xu]
	  {xu}
R.R. Xu, R. Wyss and P.M. Walker, Phys. Rev. {\bf C60} (1999) 051301

\bibitem% [tera]
          {tera}
J. Terasaki, P.-H. Heenen, P. Bonche, J. Dobaczewski and H. Flocard, 
Nucl. Phys. {\bf A593} (1995) 1

\bibitem% [rigol]
	  {rigol}
C. Rigollet, P. Bonche, H. Flocard and P.-H. Heenen, Phys. Rev. C59 (1999) 3120

\bibitem% [fallon]
	  {fallon} 
P. Fallon, P.-H. Heenen, W. Satula, R.M. Clark, F.S. Stephens, M.A. Deleplanque, R.M. Diamond, I.Y. Lee, A.O. Macchiavelli and K. Vetter, Phys. Rev. {\bf C60} (1999) 044301

\bibitem% [cwiok1]
	  {cwiok1}
S. \'Cwiok, W. Nazarewicz and  P.-H. Heenen, Phys. Rev. Letters {\bf 83} (1999) 1108

\bibitem% [dug]
	  {dug}
T. Duguet, P. Bonche and P.-H. Heenen, Nucl. Phys. {\bf A679} (2001) 427

\bibitem% [bonche]
	  {bonche}
P. Bonche, H. Flocard, P.-H. Heenen, S.J. Krieger and M.S. Weiss, Nucl. Phys. {\bf A443} (1985) 39

\bibitem% [audi]
          {audi}
G. Audi and A.H. Wapstra, Nucl. Phys. {\bf A595} (1995) 409

\bibitem% [dech]
          {dech}
J. Decharg\'e and D. Gogny, Phys. Rev. {\bf C21} (1980) 1568

\bibitem% [doba]
	  {doba}
J. Dobaczewski, P. Magierski, W. Nazarewicz, W. Satula and Z. Szyma\'nski, Phys. Rev. {\bf C63} (2001) 024308

\bibitem% [flo]
	  {flo}
H. Flocard, Cours d'\'et\'e des Houches (2000), to be published

\bibitem% [satupol]
          {satupol}
W. Satula,  Plenary talk at Nuclear Structure'98, AIP Conference Proceedings 481, ed. C. Baktash
           American Inst. of Physics, New York 1999, p. 141

\bibitem% [rutz2]
	  {rutz2}
K. Rutz, M. Bender, P.-G. Reinhard and J.A. Maruhn, Phys. Lett. {\bf B468} (1999) 1

\bibitem% [hilaire]
	  {hilaire}
S. Hilaire, J.-F. Berger, M. Girod, W. Satula and P. Schuck, unpublished


\bibitem% [wyss1]
	  {wyss1}
W. Satula and R. Wyss, Phys. Lett. {\bf B393} (1997) 1; W. Satula and R. Wyss, LANL preprint nucl-th 0010041; ibid. 0011056, unpublished; W. Satula and R. Wyss, Invited Talk at the International Conference ``High spin 2001'', February 6-11, 2001, Warsaw, Poland

\bibitem% [vogel]
	  {vogel}
P. Vogel, Nucl. Phys. {\bf A662} (2000) 148

\bibitem% [tera2]
	  {tera2}
J.~Terasaki, R.~Wyss and P.-H.~Heenen, Phys. Lett. {\bf B437} (1998) 1




\end{thebibliography}
\end{document}